\documentclass[11pt]{amsart}
\usepackage{geometry}                
\usepackage{graphicx}
\usepackage{amssymb}
\usepackage{epstopdf}
\usepackage{array}
\usepackage{subfigure}
\usepackage{color}
\usepackage{url}

\definecolor{gray}{rgb}{0.7,0.7,0.7}

\usepackage{supertabular}
\usepackage{hhline}
\newcommand\arraybslash{\let\\\@arraycr}
\makeatother
\newtheorem{finding}{Finding}

\usepackage{lscape}

\title[Route 20, autobahn 7, and P. polycephalum]{Route 20, autobahn 7 and Physarum polycephalum:\\ Approximating longest roads in USA and Germany \\ with slime mould  on 3D terrains }
\author[Adamatzky]{Andrew Adamatzky\\ University of the West of England, Bristol, UK}
\address[Adamatzky]{University of the West of England, Bristol, UK}
\email{andrew.adamatzky@uwe.ac.uk}

\begin{document}

\maketitle

\begin{abstract}
 Acellular slime mould \emph{Physarum polycephalum} is a monstrously large single cell visible by an unaided eye. 
 It shows sophisticated behavioural traits in foraging for nutrients and 
 developing an optimal transport network of protoplasmic tubes spanning sources of nutrients. The slime mould sufficiently approximates man-made transport networks on a flat substrate.   Does slime mould imitate man-made transport networks on three-dimensional terrain as well as it does on a flat substrate? We simplified the problem to 
 approximation of a single transport route. In laboratory  experiments with 3D Nylon terrains of USA and Germany we imitated development of route 20, the longest road in USA, and  autobahn 7, the longest national motorway in Europe. We found that slime mould builds longer transport routes on 3D terrains, comparing to flat agar plates yet sufficiently approximates man-made transport routes studied.  We discuss how nutrients placed in destination sites affect performance of slime mould, how the mould navigates around elevations, and offer an automaton model of slime mould which might explain variability of the protoplasmic routes constructed.  We also display results of scoping experiments with slime mould on 3D terrains of Russia and United Kingdom.\\ 

\end{abstract}

\section{Introduction}

The increase of long-distance travel and subsequent reconfiguration of vehicular  and social networks requires novel and unconventional approaches towards analysis of dynamical processes in complex transport networks,  routing and localisation of vehicular networks, optimisation of interactions between different parts of a transport network during scheduling of the road expansion, and shaping of transport network structure. Historical development and 
reconfiguration of major transport routes  is somewhat analogous to biological growth and pattern formation. Are man-made networks really similar to biological transport networks? Can living substrates develop better, in terms of minimal distances and maximum coverage, transport networks than humans made? There is only one way to answer the questions: by undertaking laboratory experiments with a living substrate that produces transport networks.  Ants and plant roots are first instances that come to mind, however it is somewhat difficult to experiment with these creatures. In 2001 Nakagaki with colleagues published a paper on shortest path approximation with live plasmodium of \emph{Physarum polycephalum}~\cite{nakagaki_2001}. Three years later Tsuda, Aono and Gunji  implemented Boolean logical gates with live slime  mould~\cite{tsuda_2004}. Thus  the scientific world was introduced to a unique living computing substrate, which is easy to cultivate and handle, monitor and experiment with.  

Plasmodium is a vegetative stage of acellular slime mould \emph{P. polycephalum}, a single cell with many nuclei, which feeds on microscopic particles~\cite{stephenson_2000}. When foraging for its food the plasmodium propagates towards sources of food, surrounds them, secretes enzymes and digests the food; it may form a congregation of protoplasm covering the food source. When several sources of nutrients are scattered in the plasmodium's range, the plasmodium forms a network of protoplasmic tubes connecting the masses of protoplasm at the food sources. A structure of the protoplasmic networks is apparently optimal, in a sense that it covers all sources of nutrients and provides a robust and speedy transportation of nutrients and metabolites in the 
plasmodium's body~\cite{nakagaki_2001,nakagaki_2001a,nakagaki_iima_2007}.  

To uncover analogies between biological and human-made transport networks and to project behavioural traits of biological networks onto development of vehicular transport networks we conducted a series of experimental laboratory studies on evaluation and approximation of motorway networks by \emph{P. polycephalum} in fourteen geographical regions:   Africa, Australia, Belgium, Brazil, Canada, China, Germany, Iberia, Italy, Malaysia, Mexico, The Netherlands, UK, and USA~\cite{adamatzky_bioevaluation}. We represented each region with an agar plate, imitated major urban areas with oat flakes, inoculated plasmodium of \emph{P. polycephalum} in a capital, and analysed structures of protoplasmic networks developed. For all regions studied~\cite{adamatzky_bioevaluation}, 
 networks of protoplasmic tubes grown by plasmodium match, at least partly, the networks of man-made transport routes. The shape of a country and the exact spatial distribution of urban areas, represented by sources of nutrients, may play a key role in determining the exact structure of the plasmodium network. The experiments~\cite{adamatzky_bioevaluation} had one obvious deficiency: slime mould propagated on a flat substrates, agar plates. Poor matching between protoplasmic networks and man-made transport networks in some countries could have been attributed to particular configurations of elevations and depressions in the countries's terrains. How does slime mould \emph{P. polycephalum} behave on a 3D terrain? 
 
 We envisaged that plasmodium of \emph{P. polycephalum} will not completely 'ignore' 3D topography because the slime mould is gravisensitive and positively geotropic~\cite{block_1986,block_1998}.  Moreover, plasmodium shows morphological geopolarity, as discovered in~\cite{block_1989}: the ectoplasmic wall of a slime mould tube, lying or hanging on horizontal surface, is much thinner on side closer to earth.  We expected that being placed on a 3D terrain with a source of nutrients slime mould would propagate towards the source of nutrients and navigate around elevations due to positive geotropism and relatively lower humidity of the elevations.

Does slime mould on a 3D terrain match man-made transport routes better or worse than on a flat substrate? 
We conducted experiments on approximation of route 20 in USA and autobahn 7 in Germany with slime mould inoculated on a 3D plastic terrains. Why these specific routes?  Route 20  (Fig.~\ref{schemesroutes}a) is the longest highway in USA, 5,415 km, it starts in Boston, Massachusetts, and end in Newport, Oregon (Fig.~\ref{schemesroutes}a)~\cite{nelson_2008,lewis_1997,mcnichol_2005,karnes_2009}. Route 20 crosses USA from east coast to west coast. Autobahn 7  (Fig.~\ref{schemesroutes}a) 963 km is the longest national motorway in Europe. It starts in Flensburg, passes via Hamburg, Hannover, Kassel, Fulda, W\"{u}rzburg, Ulm, Kempten and end in F\"{u}ssen (Fig.~\ref{schemesroutes}b)~\cite{brilon_1994, zeller_2007}.
 
 The paper is structured as follows. We described experimental setup in Sect.~\ref{methods}. Section~\ref{results} presents results of laboratory experiments of slime mould of 3D terrains and their analysis. Using excitable cellular automata models we imitate variability of slime mould routes in Sect.~\ref{simulation}. In Sect.~\ref{UKRussia} we test our approach to slime mould path based path finding on terrains of United Kingdom and Russia. A summary is provided in Sect.~\ref{discussion}.

\section{Methods}
\label{methods}

\begin{figure}[!tbp]
\centering
\subfigure[]{\includegraphics[width=0.58\textwidth]{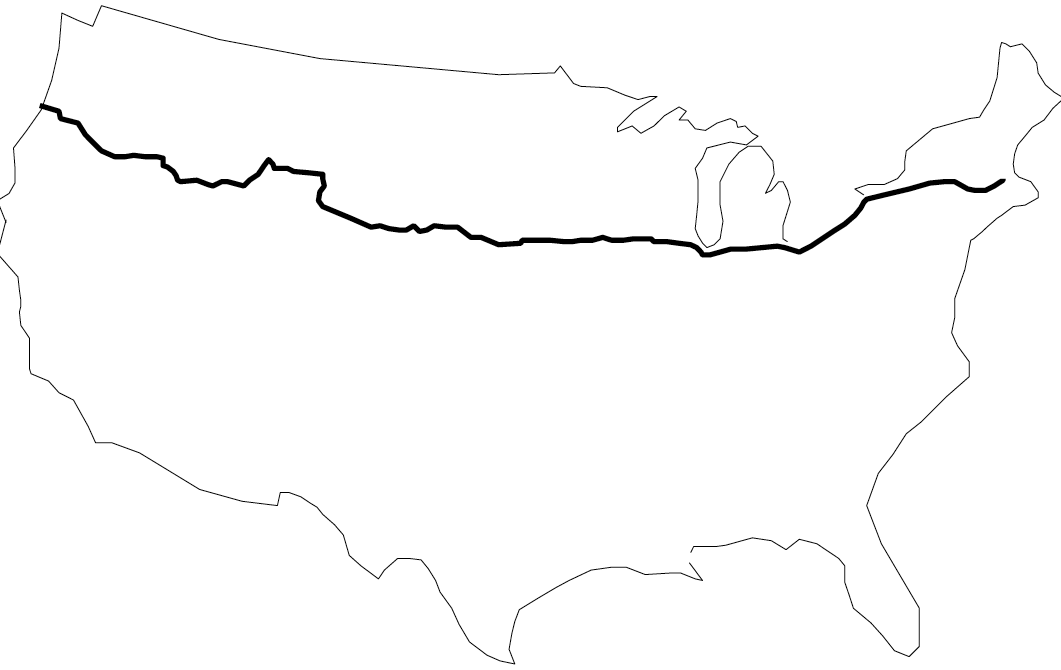}}
\subfigure[]{\includegraphics[width=0.38\textwidth]{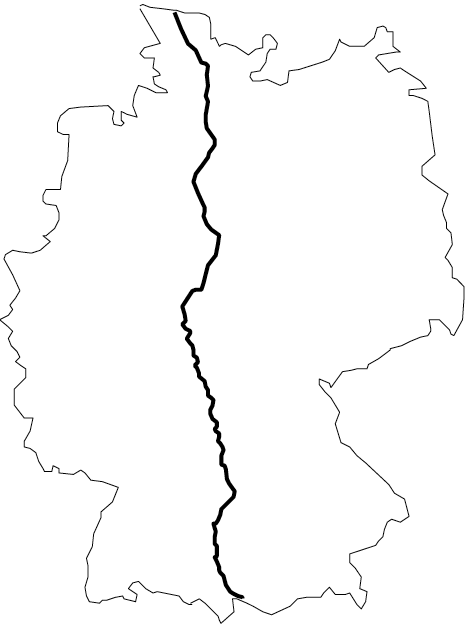}}
\subfigure[]{\includegraphics[width=0.9\textwidth]{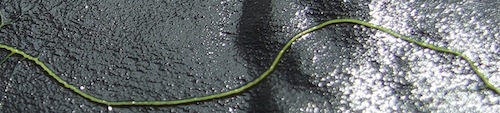}}
\caption{Schemes of route 20 (a) and autobahn 7 (b).  (c)~A protoplasmic tube of \emph{P. polycephalum} on a bare Nylon substrate.}
\label{schemesroutes}
\end{figure}

\begin{figure}[!tbp]
\centering
\subfigure[]{\includegraphics[width=0.49\textwidth]{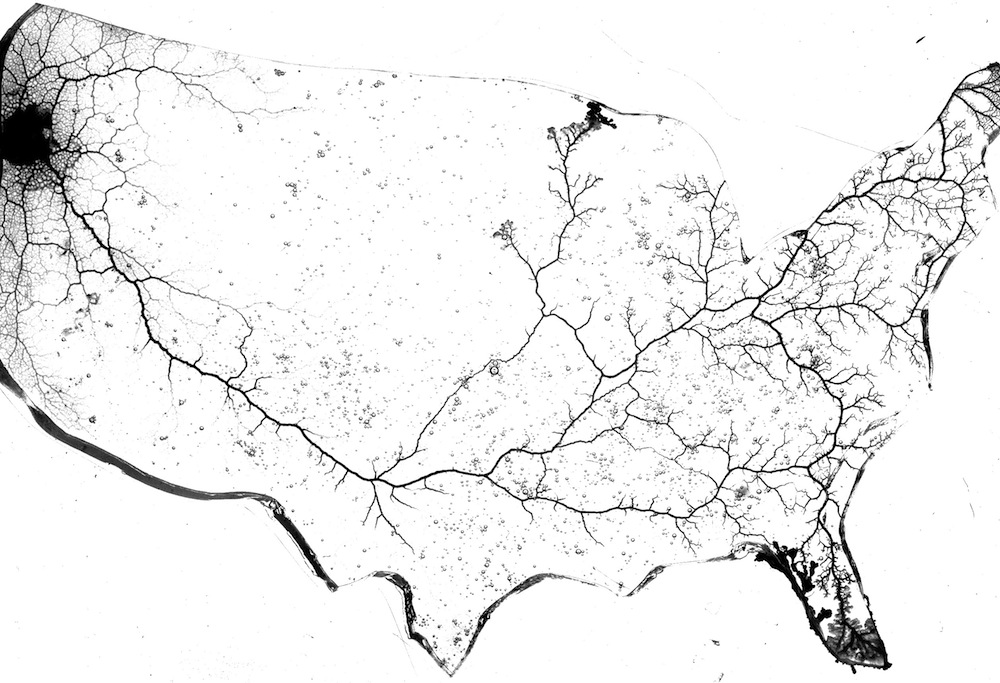}}
\subfigure[]{\includegraphics[width=0.49\textwidth]{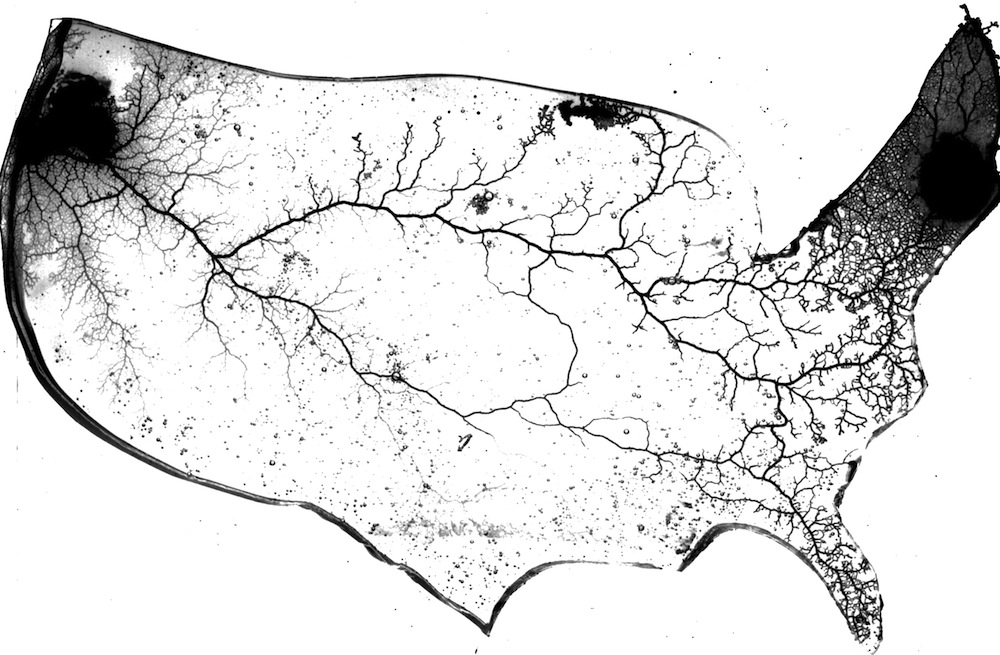}}
\subfigure[]{\includegraphics[width=0.85\textwidth]{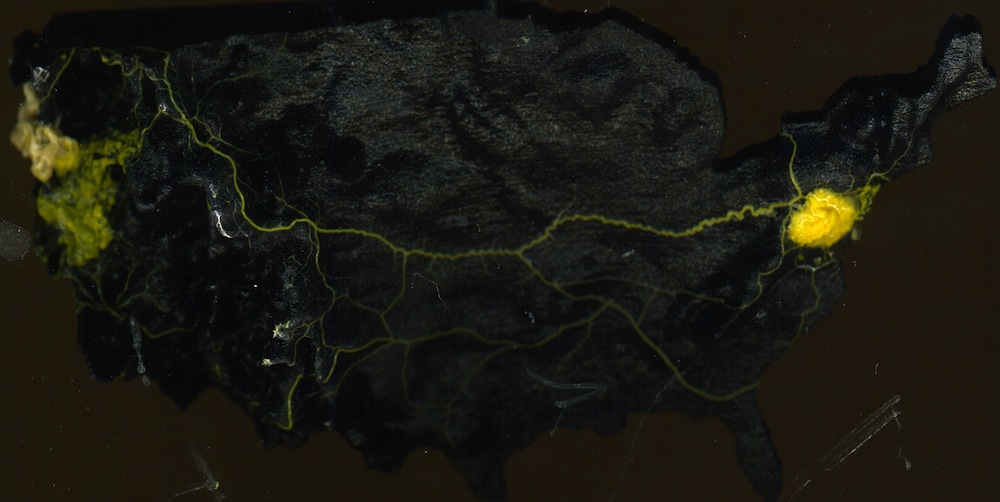}}
\caption{Examples of slime mould routes developed on (a)~USA shaped agar plate without attractants, 
(b)~USA shaped agar plate with an oat flake placed in Boston, (c)~3D terrain of USA with an oat flake placed in Boston. }
\label{USAExamples}
\end{figure}

\begin{figure}[!tbp]
\centering
\subfigure[]{\includegraphics[width=0.44\textwidth]{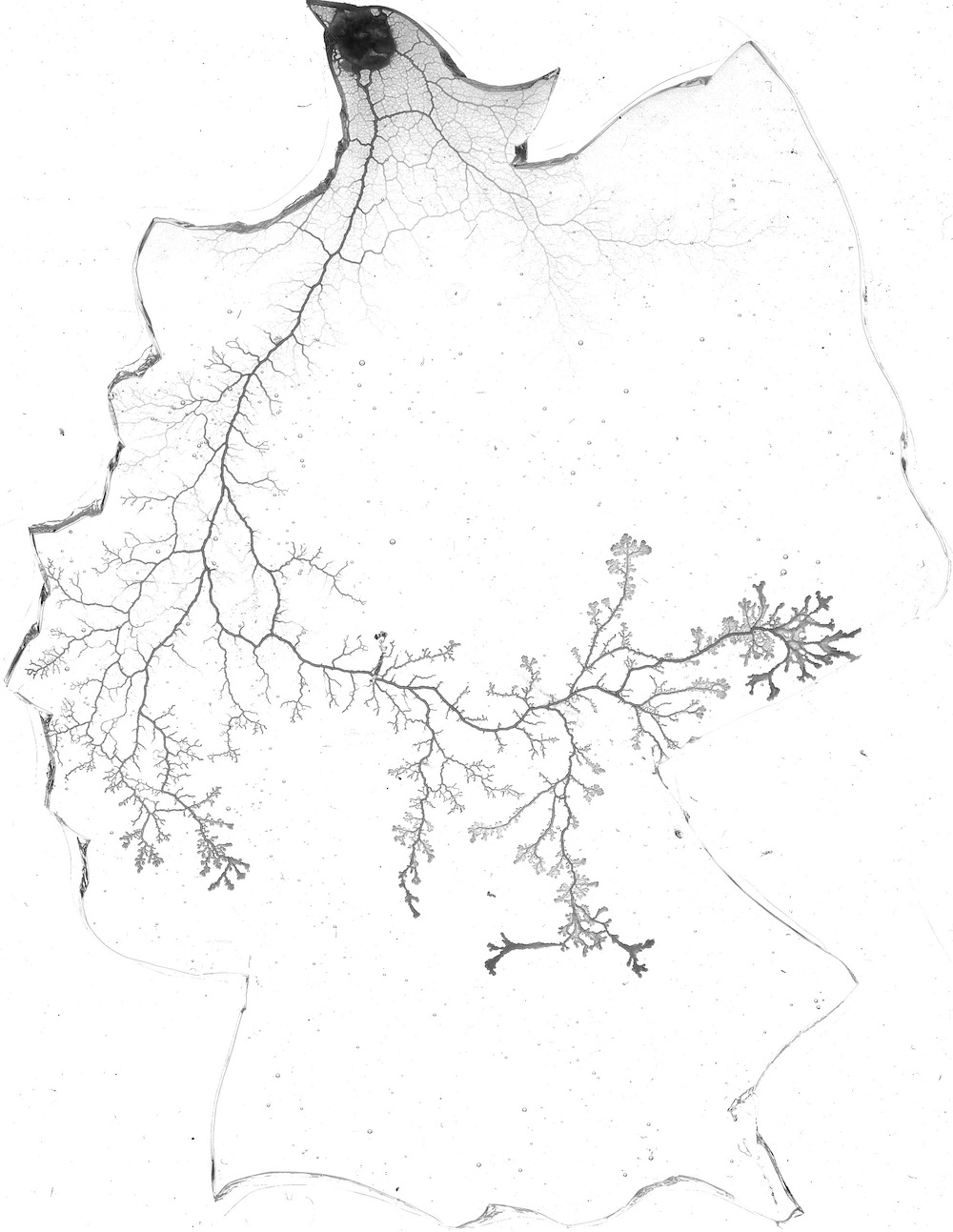}}
\subfigure[]{\includegraphics[width=0.44\textwidth]{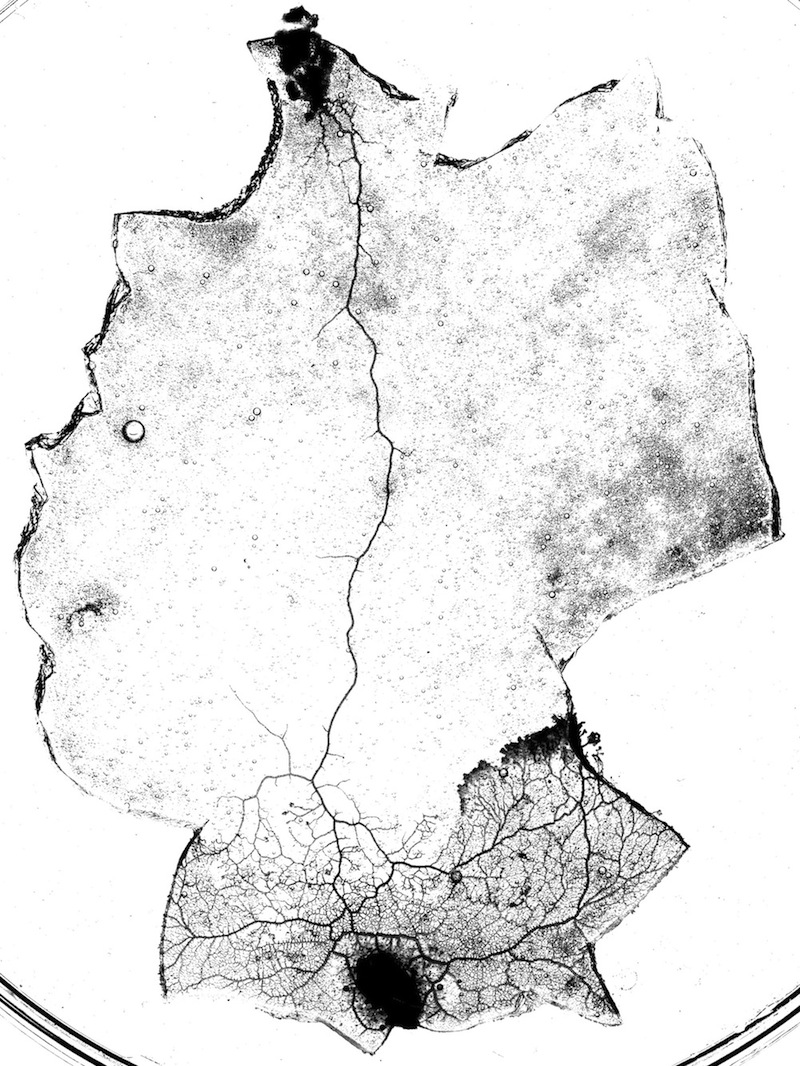}}
\subfigure[]{\includegraphics[width=0.49\textwidth]{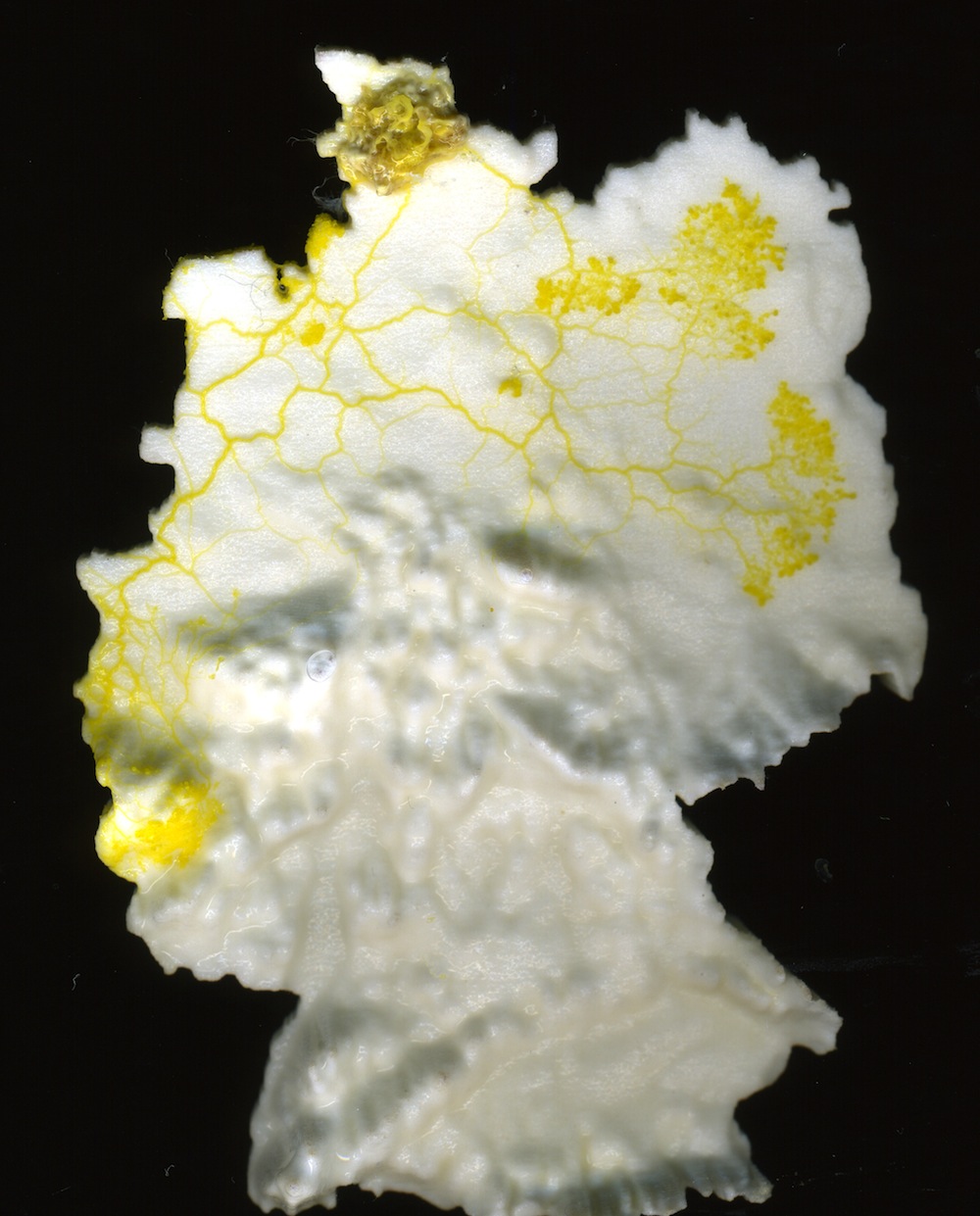}}
\subfigure[]{\includegraphics[width=0.38\textwidth]{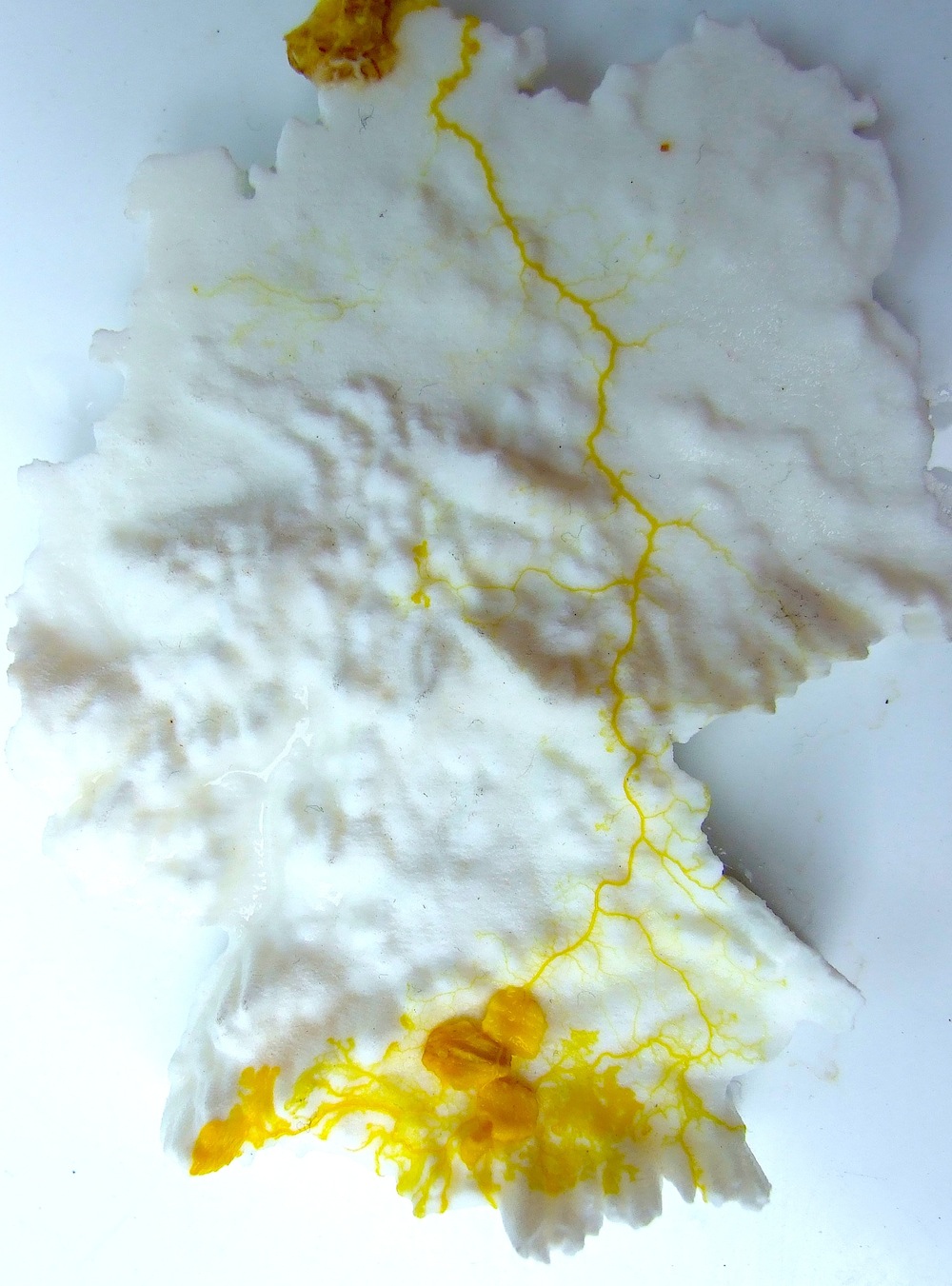}}
\caption{Examples of slime mould routes developed on (a)~Germany shaped agar plate without attractants, 
(b)~Germany shaped agar plate with an oat flake placed in F\"{u}ssen, (c)~3D terrain of Germany without attractants in F\"{u}ssen,  and (d)~3D terrain of Germany with an oat flake placed in F\"{u}ssen. }
\label{GermanyExamples}
\end{figure}

3D terrains of USA and Germany were ordered in \url{http://www.printablegeography.com/}, where there were produced as follows.  The elevation data are downloaded from DIVA-GIS (\url{http://www.diva-gis.org/gdata}), 
original source is CGIAR (\url{http://srtm.csi.cgiar.org/}). The data are projected with Mercator. Each terrain was
printed using  Selective Laser Sintered PA 2200 with Nylon 12. Terrain of Germany was 8.2~cm wide, 11~cm long, and 3~cm high (dimensions measured between extremeties), and terrain of USA was 11~cm wide, 5.9~cm long and 3~cm high.

In each experiment we inoculated plasmodium of \emph{P. polycephalum} in the positions of Newport, Oregon, USA (Fig.~\ref{schemesroutes}a) and Flensburg, Germany  (Fig.~\ref{schemesroutes}b), and  oat flakes were placed in the positions of Boston, Massachusetts, USA, and F\"{u}ssen, Germany.  The terrains were 
kept in closed yet naturally ventilated containers. The terrains were not artificially 
wetted or covered with any water retaining substrate  but stayed in containers filled with water (terrains were positioned above the water, without water even touching edges of the terrains). Humidity is proved to be sufficient for the slime mould to propagate on a bare nylon surface  (Fig.~\ref{schemesroutes}c). The terrains were kept in darkness, at temperature 22-27$^\text{o}$C, except for observation and image recording. Configurations of plasmodium networks were photographed with FujiFilm FinePix 6000 camera and scanned with an Epson Perfection 4490 scanner. Experimental setups with 3D terrains are illustrated in Figs.~\ref{GermanyExamples}cd and \ref{USAExamples}c.

To understand influence of elevations on the slime mould's propagation we also conducted experiments on a flat agar gel. For these experiments we used $220 \times 220$~mm polystyrene square Petri dishes and 2\% agar gel (Select agar, by Sigma Aldrich) as a substrate. Agar plates, about 2-3~mm in depth, were cut in  shape of USA and Germany
(Figs.~\ref{GermanyExamples}ab and  \ref{USAExamples}ab).

We have conducted 11 experiments with USA-shaped flat agar, 14 experiments with 3D terrain of USA,  13 experiments with Germany-shaped flat agar, and 16 experiments with 3D terrain of Germany.

\section{Results}
\label{results}

Usually it takes slime mould 4-5 days to cover the distance between Newport and Boston and 2-3 days to propagate from  Flensburg to F\"{u}ssen (such difference in propagation time on the 3D terrains of almost the same size is possibly due to high amount of steep elevations in the  West of USA, which prevent the slime mould from gaining speed and momentum when starting colonisation of the terrain). In experiments with  Germany one day of slime mould's propagation roughly corresponds to 3-5 hours of real-life driving along the autobahn 7, and in experiments with USA to about 10 hours driving along route 20.

\begin{figure}
\centering
\subfigure[]{\includegraphics[width=0.61\textwidth]{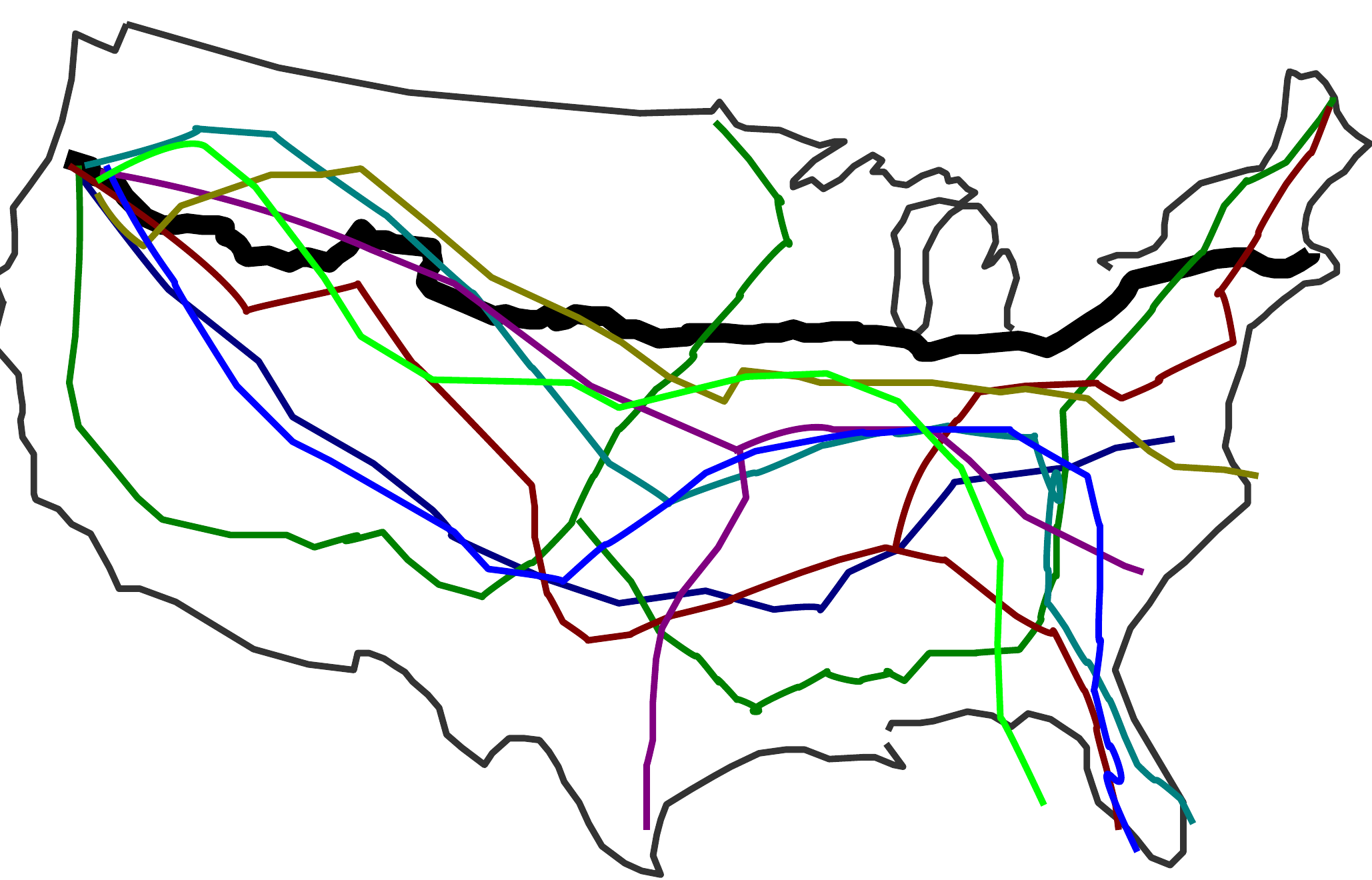}}
\subfigure[]{\includegraphics[width=0.38\textwidth]{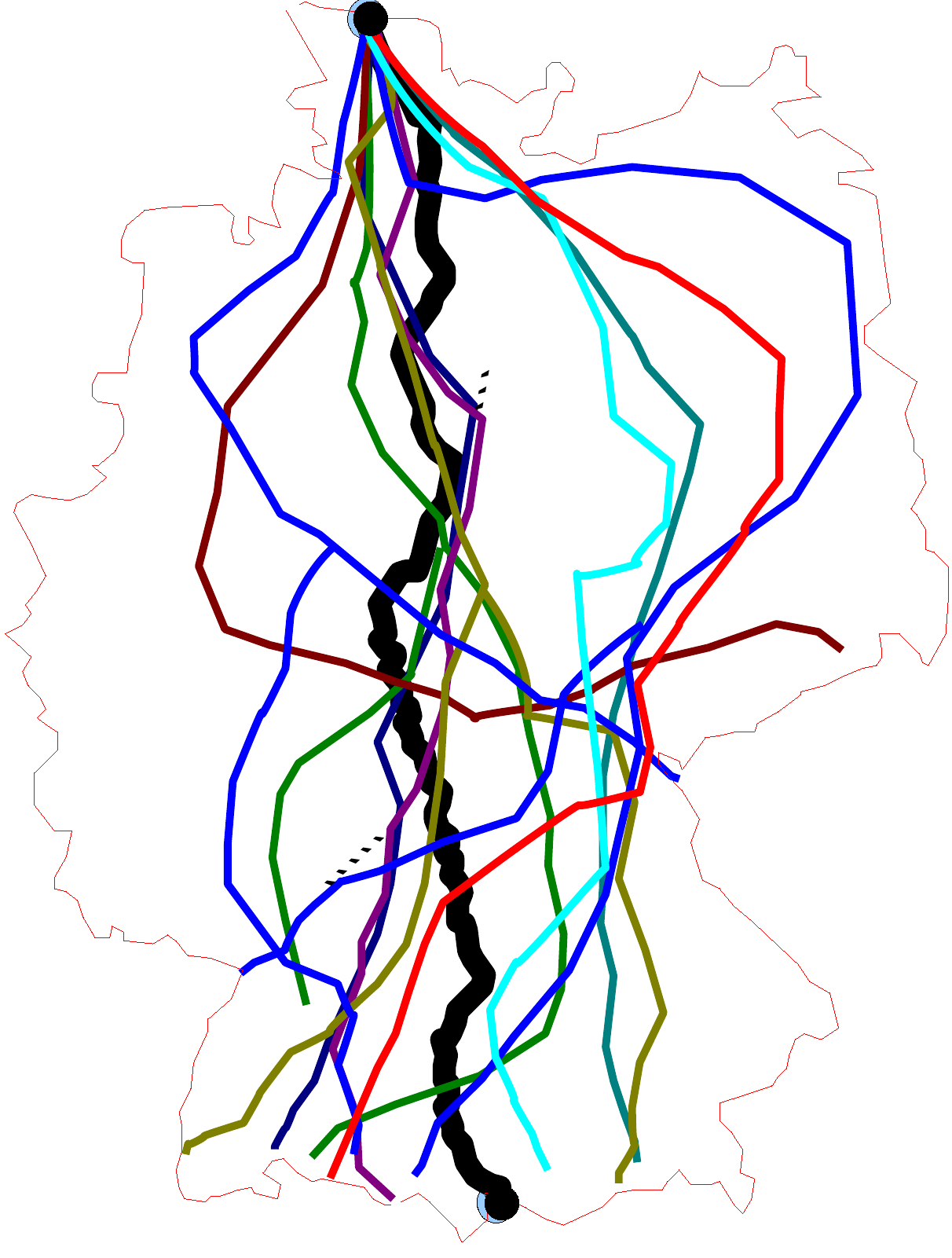}}
\subfigure[]{
\begin{tabular}{ccc}
Experiment	&	Mishits USA	&	Mishits Germany	\\ \hline \hline
1	&	0.187	&	0.072	\\
2	&	0.189	&	0.094	\\
3	&	0.276	&	0.138	\\
4	&	0.268	&	0.156	\\
5	&	0.426	&	0.192	\\
6	&	0.674	&	0.249	\\
7	&	0.726	&	0.045	\\
8	&	0.714	&	0.091	\\
9	&	0.659	&	0.112	\\
10	&	1.043	&	0.509	\\ \hline \hline
	&		&		\\
Min	&	0.187	&	0.045	\\
Max	&	1.043	&	0.509	\\
Average	&	0.516	&	0.166	\\
Median	&	0.543	&	0.125	\\
Standard deviation	&	0.289	&	0.135	\\
\end{tabular}
}
\label{schemenoattractants}
\caption{Results of experiments on growing slime mould without attractants, i.e. no oat flakes placed in 
destination sites F\"{u}ssen and Boston. (a)~USA. (b)~Germany. Major protoplasmic tubes extracted from different experiments are shown by different colours (gradations of grey). (c)~Statistics of mishits: deviations from destination sites divided by a heigh, USA, or width, Germany, of the country.}
\end{figure}

\begin{finding}
Slime mould rarely reaches a destination which is not marked by a source of nutrients.
\end{finding}

In experiments without chemoattractants the plasmodium was inoculated in start points, Flensburg and Newport, and no 
oat flakes were placed in destination sites, F\"{u}ssen and Boston. Due to proximal positions of start points the plasmodium 
was only able to propagate south and east in  USA (Fig.~\ref{USAExamples}a), and south, east and west in 
Germany (Fig.~\ref{GermanyExamples}ac).  Schemes of the major protoplasmic tubes extracted from images of ten experiments on flat agar plates are shown in Fig.~\ref{schemenoattractants}. The schemes demonstrate that slime mould's propagation was mainly limited by geometric shape of agar plate. There are clear evidences of the plasmodium 
reflection from north and south boundaries of USA, and east and west boundaries of Germany. Such interaction with the boundaries of agar plates and randomised foraging behaviour of plasmodium cause the slime mould to reach east cost
of USA (Fig.~\ref{schemenoattractants}a) and south boundary of Germany (Fig.~\ref{schemenoattractants}b) in arbitrary sites.  Mishits --- distances of plasmodium's hits from destination sites divided by  height of a country, in USA, or width in Germany --- are tabulated in Fig.~\ref{schemenoattractants}c. 

\begin{figure}[!tbp]
\centering
\subfigure[]{\includegraphics[width=0.61\textwidth]{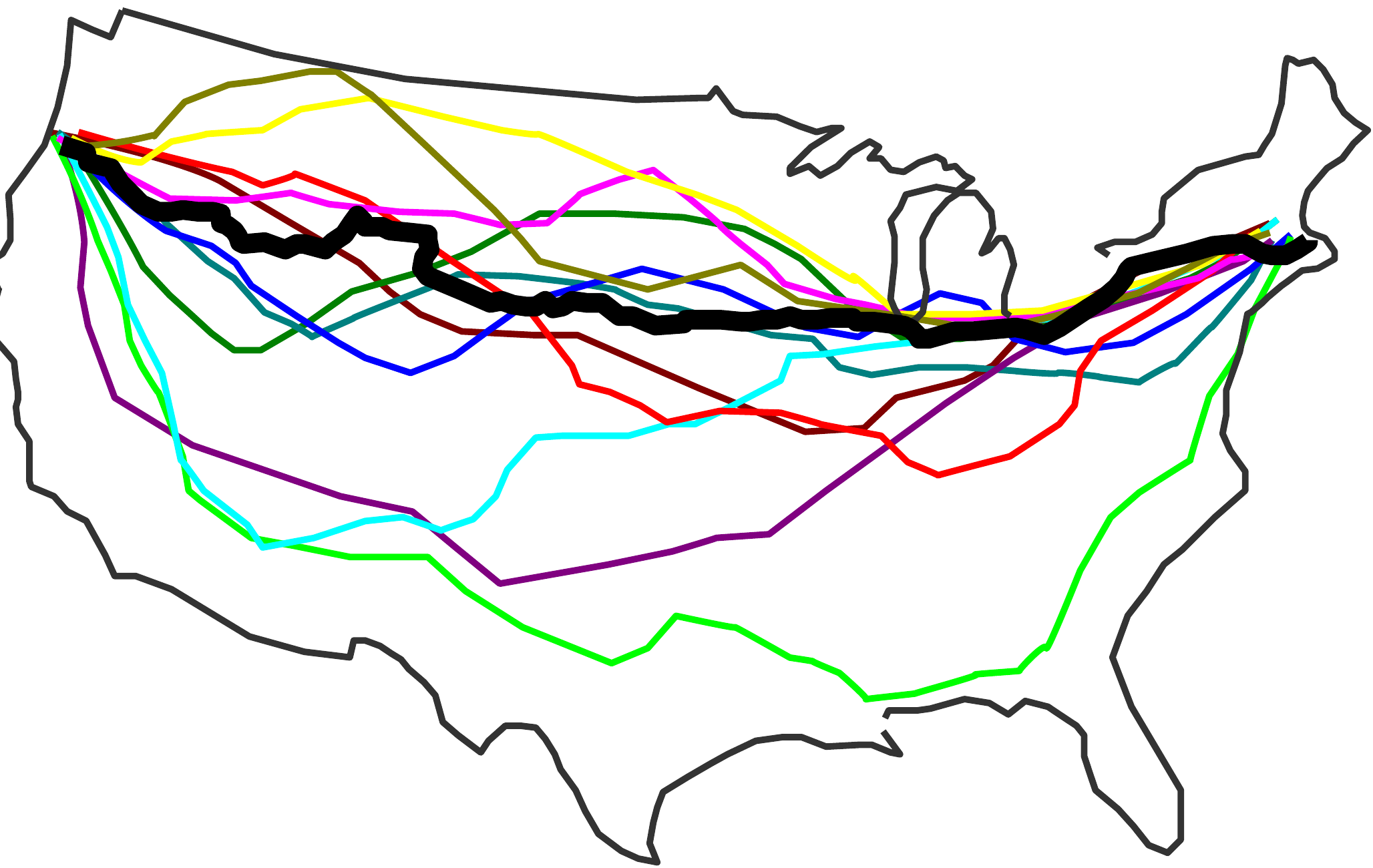}}
\subfigure[]{\includegraphics[width=0.38\textwidth]{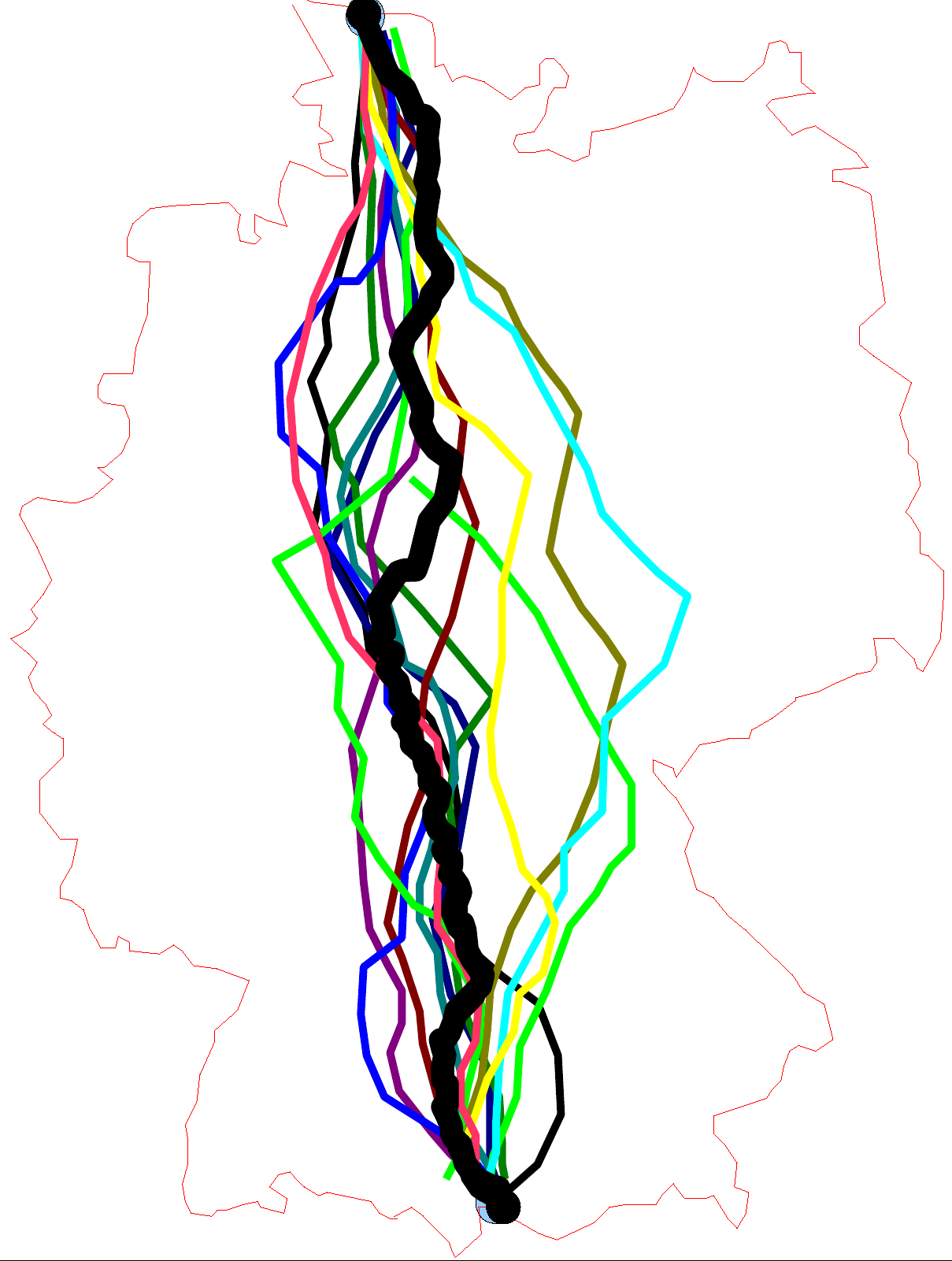}}
\subfigure[]{\includegraphics[width=0.61\textwidth]{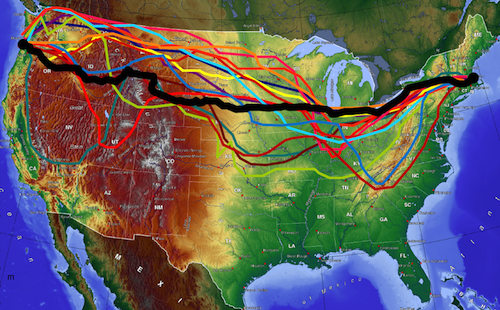}}
\subfigure[]{\includegraphics[width=0.38\textwidth]{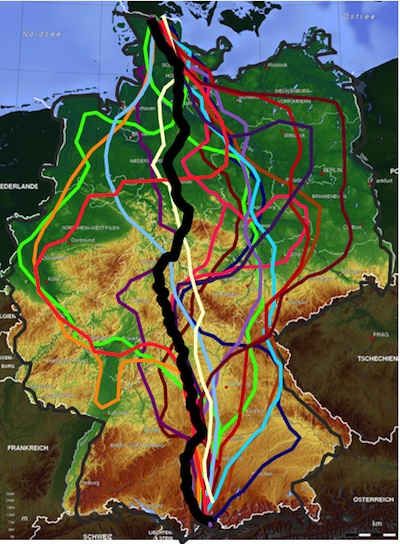}}
\caption{Results of experiments on approximation of route 20 in USA (ab) and autobahn 7 in Germany (cd) 
with slime mould. Slime mould is inoculated in Newport, Oregon (ab) and Flensburg (cd) and oat flakes are 
placed in Boston, Massachusetts (ab) and F\"{u}ssen (cd).  Experiments are conducted on flat agar plates (ac) and 3D Nylon terrains (bd). Major protoplasmic tubes extracted from different experiments are shown by different colours (gradations of grey).}
\label{USAGermanyResults}
\end{figure}

\begin{table}[!tbp]
\caption{Lengths of slime mould routes from start sites to destination sites divided by length of route 20 in USA and length of autobahn 7 in Germany. }
\begin{tabular}{ccccc}
Experiment	&	USA Agar	&	USA 3D	&	Germany Agar	&	Germany 3D	\\ \hline \hline
1	&	0.963	&	1.049	&	1.005	&	1.290	\\
2	&	1.028	&	1.056	&	1.030	&	1.117	\\
3	&	0.924	&	0.979	&	0.962	&	1.280	\\
4	&	0.965	&	1.278	&	0.964	&	1.010	\\
5	&	0.954	&	0.984	&	1.016	&	1.130	\\
6	&	1.020	&	0.956	&	1.048	&	1.252	\\
7	&	1.009	&	1.028	&	1.016	&	1.037	\\
8	&	1.026	&	1.108	&	0.981	&	1.065	\\
9	&	1.147	&	1.110	&	0.994	&	1.039	\\
10	&	1.325	&	1.156	&	0.971	&	1.093	\\
11	&	1.148	&	1.342	&	1.014	&	0.995	\\
12	&		&	1.211	&	1.093	&	1.413	\\
13	&		&	1.069	&	0.967	&	1.077	\\
14	&		&	0.997	&		&	1.040	\\
15	&		&		&		&	1.359	\\
16	&		&		&		&	1.327	\\
	&		&		&		&		\\\\ \hline \hline
	&		&		&		&		\\
Min	&	0.924	&	0.956	&	0.962	&	0.995	\\
Max	&	1.325	&	1.342	&	1.093	&	1.413	\\
Average&	1.046	&	1.095	&	1.005	&	1.158	\\
Median	&	1.020	&	1.063	&	1.005	&	1.105	\\
Standard deviation	&	0.117	&	0.116	&	0.038	&	0.139	\\
\end{tabular}
\label{tablestatistics}
\end{table}

\begin{figure}[!tbp]
\centering
\subfigure[]{\includegraphics[width=0.8\textwidth]{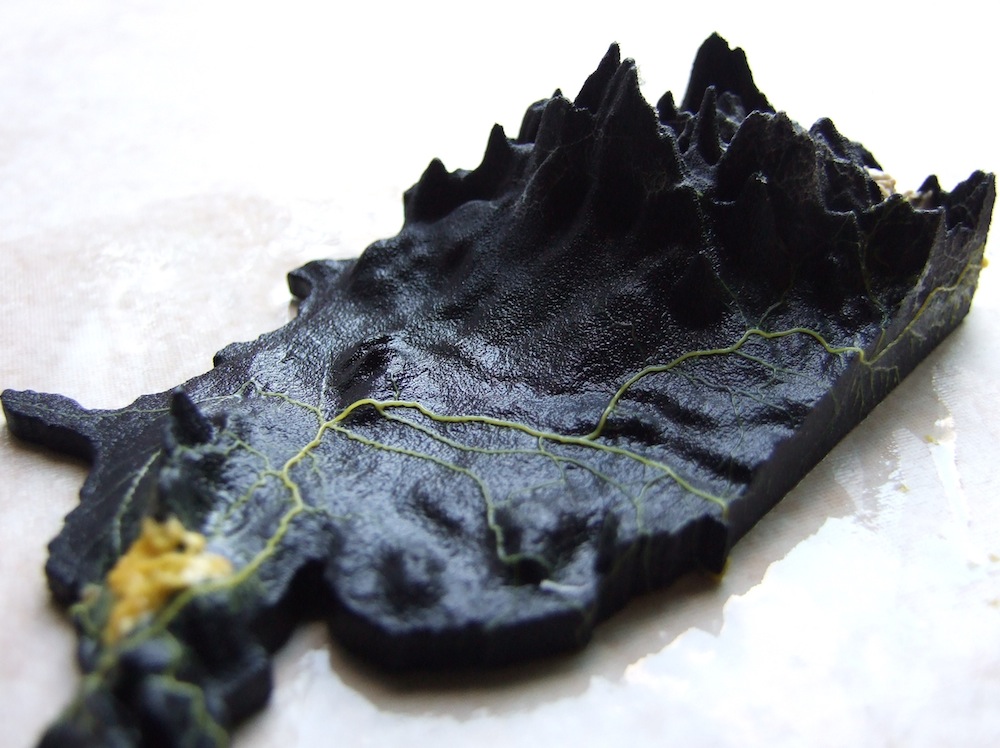}}
\subfigure[]{\includegraphics[width=0.8\textwidth]{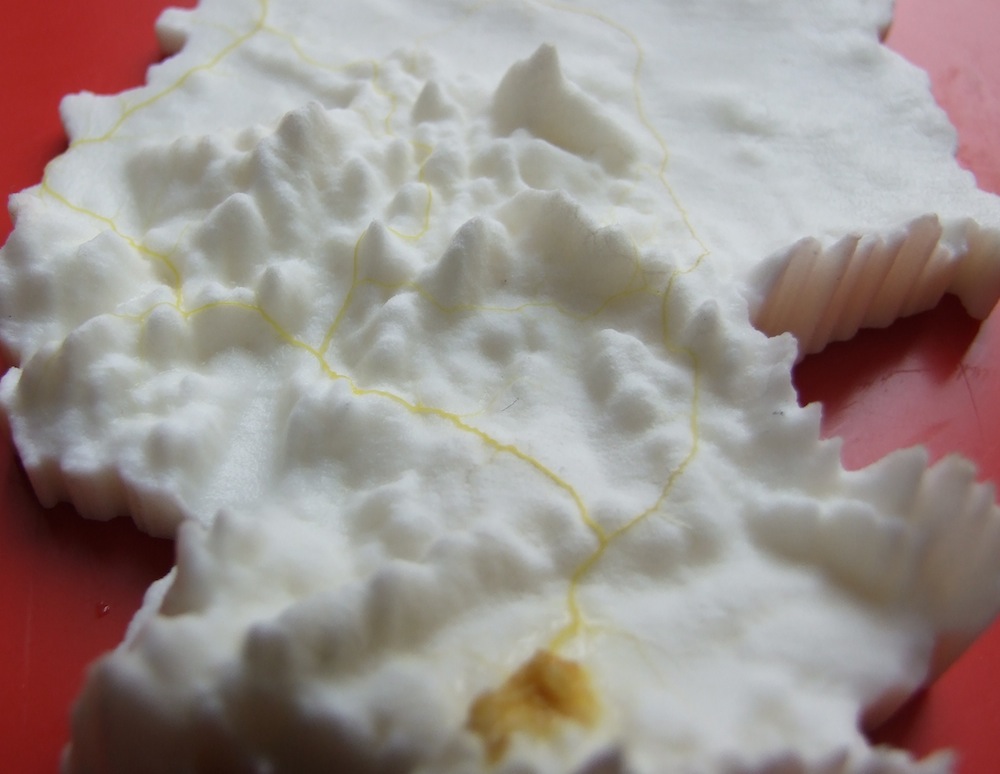}}
\caption{Examples of slime mould navigating around mountains in USA (a) and Germany (b).}
\label{USAGermanyMountains}
\end{figure}

\begin{finding}
Slime mould approximates autobahn 7 in Germany 1.041 times better than it approximates route 20 in USA on agar plates.
\end{finding}

Slime mould's behaviour on flat agar plates with attractants, oat flakes placed in Boston and F\"{u}ssen was, rather partially, determined by chemo-attractants diffusing from oat flakes (Figs.~\ref{USAExamples}b and~\ref{GermanyExamples}b).  The plasmodium demonstrated less foraging activity towards south in  
USA (Fig.~\ref{USAGermanyResults}a) and even less towards east and west in Germany 
(Fig.~\ref{USAGermanyResults}b). Transport links built by slime mould between Newport and Boston are 1.046 times longer in average than a length of route 20 and the slime mould transport links between 
Flensburg and F\"{u}ssen are just 1.005 times longer in average than a length of autobahn 7  
(Tab.~\ref{tablestatistics}). Assuming that chemo-attractants are not only airborne but also diffuse in a substrate, agar gel, shape of Germany provides favourable conditions for slime mould to detect diffusing chemoattractants: in USA the source of attractants is masked by indentations made by Lake Ontario, Quarry, Lake  Huron and Lake Michigan (Fig.~\ref{USAGermanyResults}a). This may lead to prevalent diffusion of chemoattractants southwards and subsequent deviation of growing slime mould towards southern parts of USA.

\begin{finding}
Slime mould approximates  route 20 in USA 1.095 times better than it approximates autobahn 7 in Germany on 3D terrains. 
\end{finding}

Table~\ref{tablestatistics} shows statistics on lengths of major tubes representing route 20 and autobahn 7 on 3D terrains. On average, slime mould routes are 1.095 times longer than route 20, and 
1.158 longer than autobahn 7.  In experiments with 3D terrains slime mould grows on  bare Nylon terrains. 
Thus even if diffusion of chemoattractants occurred in the thin film of water covering the terrains, due to high 
humidity in closed containers, we believe the airborne attractants would play a major role in guiding the 
plasmodium toward destinations. Absence of any guidance force in a substrate together with non-flat terrain 
(Fig.~\ref{USAGermanyMountains}) cause the slime mould to deviate, sometimes substantially, from a shortest route towards its destination site (Figs.~\ref{USAExamples}c and \ref{GermanyExamples}d). 

In experiments with 3D terrain of USA slime mould is inoculated in Newport, Oregon. As soon as the plasmodium recovers from inoculation procedure and begins exploration of a surrounding environment it finds that its way towards
east is blocked by Cascade Range mountains.

In six out of fourteen experiments with 3D terrain of USA slime mould (Fig.~\ref{USAGermanyResults}c) propagates north along Cascade Range mountains. If the 3D terrain included Canada the plasmodium would propagate further north, however in present experiments the slime mould turns east along national boundary of USA. It crosses Rocky Mountains at the sites of Glacier National Park and 
Flathead National Forest. The plasmodium then passes through North and South Dakota. In Minnesota plasmodia (recorded in different experiments) split: half of them continues into Wisconsin along interstate 94, another half turns south to Iowa, along interstate 45, and then turn east to Illinois along interstate 70. On encountering Lake Michigan the 'Winsconsin bunch' of plasmodia also moves to Illinois.  Scenario of further propagation is very common and registered in nine out of twelve experiments: the slime mould propagates through Indiana, Ohio, Pennsylvania, New York and, finally, enter the its destination site in  Massachusetts  (Fig.~\ref{USAGermanyResults}c). 

In seven out of fourteen experiments with 3D terrain of USA slime mould propagates along bed of Columbia river and 
crosses Cascade Range between Gifford Pinchot and Mountain Hood National Forests (Fig.~\ref{USAGermanyResults}c).
The plasmodium crosses Rocky Mountains range north of Yellowstone National Park, in the area between Butte and Bozeman. After traversing Rocky Mountains protoplasmic routes deviate: in different experiments slime mould can pass through Wyoming, Iowa, Minnesota. 

Only in one experiment we observed  initially propagating along west coast of USA, west of Cascade Range, and along
Sierra Nevada, parallel to interstate 5. The slime mould then turns east in region between Los Angeles and Las Vegas, move north-east, propagates along interstate 15. The slime mould passes nearby Salt Lake and in the region of Butte and Bozeman joins routes adopted by majority of slime moulds. 

On approaching Boston, most slime mould routes tightly match route 20, and only two protoplasmic tubes deviate east. 
One tube follows interstate 81 and another propagates along interstate 85. 

In contrast to USA, slime mould propagating on 3D terrain of Germany shows much higher variability in configurations of major protoplasmic routes (Fig.~\ref{USAGermanyResults}d). On leaving Schleswig-Holstein slime moulds's trajectories 
dissipate west following autobahn 22 (five out of sixteen experiments) and east (ten our of sixteen experiments). 
Two of the eastern routes propagate along autobahn 20, while others positions themselves between autobahn 7 and autobahn 26. Only in one experiment slime closely followed autobahn 7 from the inoculation site. 

The western routes propagate along autobahns 36, 37 and 3, and one protoplasmic tube makes a detour south along autobahn 5 and then returning back north. Three out of five western routes, developed by slime mould,
pass in the areas of Osnabr\"{u}k, M\"{u}nster, Bonn, Mainz and join autobahn 7 in the region between Crailsheim and Heidenheim. Two western routes join autobahn 7 in the region between Hildesheim and Kassel.  Protoplasmic tubes 
spreading in eastern Germany join autobahn 7 either in the region of W\"{u}zbourg or F\"{u}ssen (Fig.~\ref{USAGermanyResults}d).

\begin{finding}
Transport links developed by slime mould of \emph{P. polycephalum} are longer than route 20 and autobahn 7. 
\end{finding}

How often do slime mould build routes shorter than route 20 and autobahn 7. In Tab.~\ref{tablestatistics} we find that 
slime mould propagating in USA  builds transport links shorter than route 20 in circa 36\% of experiments on flat agar and around 29\% of experiments on 3D terrain. Slime mould growing on Germany-shaped flat agar develops protoplasmic tubes shorter than autobahn 7 in nearly 46\% of experiments, while on 3D terrain in just one of fifteen,
7\%, experiments.   Possible explanations would be that road engineers were allowed to dig tunnels through mountains and built bridges while slime mould did not do that.  

\section{Understanding slime mould via computer simulation}
\label{simulation}

Living systems are high variable, they rarely behave exactly the same way again and again even in rigid experimental setups. \emph{P. polycephalum} does not make an exclusion: no two paths between start and destination sites are the same. Is it because of intrinsic randomness of slime mould or for some other reasons? We believe  exact routes of plasmodium propagation are determined by an agility of plasmodium, they depend on the plasmodium's physiological state and level of metabolism. Active and vigour slime mould can climb up and cross elevations. Thus it undertakes less deviations and travels towards its destination site more close to as-crow-flies path. We show in computer simulation that indeed topology of protoplasmic routes depends on a level of plasmodium's activity. We represent 'activity' as a level of excitation in a two-dimensional cellular automaton.

\begin{figure}[!tbp]
\centering
\subfigure[$\epsilon=30$]{\includegraphics[width=0.48\textwidth]{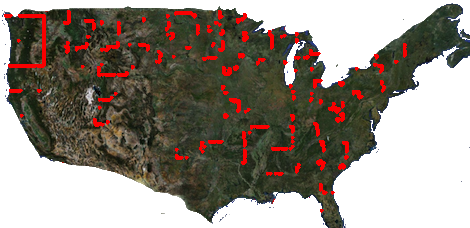}}
\subfigure[$\epsilon=30$]{\includegraphics[width=0.48\textwidth]{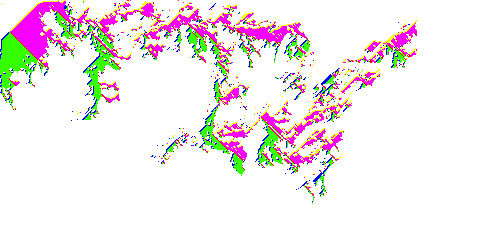}}
\subfigure[$\epsilon=150$]{\includegraphics[width=0.48\textwidth]{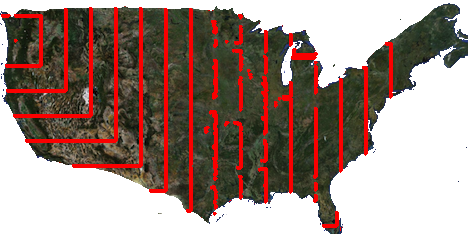}}
\subfigure[$\epsilon=150$]{\includegraphics[width=0.48\textwidth]{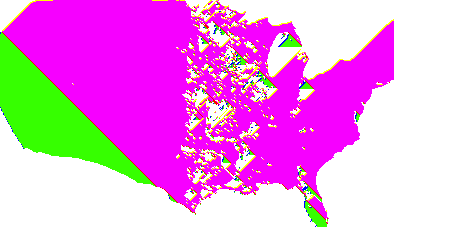}}
\caption{(a)~Time-lapse configuration of excitation wave-fronts propagating from Newport to Boston on a satellite image of USA. The wave-fronts are displayed at every 25th step of iteration. (b)~Configurations of pointers imposed by the wave-fronts in (a). Eight major orientations are represented by unique colours (gradations of grey). 
$\theta=150$,  $d=10$, $\delta \theta=10$. Cellular automaton has 500$\times$253 cells.}
\label{CA_US_POINTERS}
\end{figure}
 
 There are many ways to imitate growth and behaviour of \emph{P. polycephalum}, 
 including flow networks~\cite{tero_2008}, numerical integration of partial differential equations~\cite{adamatzky_2009}, multi-agent systems~\cite{jones_2011} and their combinations with cellular 
 automata~\cite{gunji_2011,shirakawa_2012}.  Based on our previous successful experience of imitating 
 slime mould with localised wave-fronts in Oregonator model~\cite{adamatzky_2009} we decided to imitate
 growth of the plasmodium with excitation wave-fronts in cellular automaton and extract shortest path using pointers. Such approach on approximation of a shortest path in cellular automata~\cite{adamatzky_1996}
works well not only in cellular automaton models but in crystallisation based computing devices~\cite{adamatzky_2011}.

We imitate active zones of growing slime mould in a 2D cellular automaton. The automaton is a rectangular 
 array $\mathbf A$ of cells, finite state machines, which take four states: resting $\circ$, excited $+$, refractory $-$ and precipitate $\#$. Each cell $x$ receives  RGB values $(\rho_x, \beta_x, \gamma_x)$ of the corresponding pixel $x$
 of a satellite image of the country under study (satellite images are taken in \url{maps.google.com} and reduced to sizes 500$\times$253 pixels, USA, and 505$\times$707, Germany). Let $x^t$ be a state of cell $x$ at time step $t$. All cells update their states
 simultaneously by the same rule, depending on states of their neighbours $u(x)=\{ y \in \mathbf{A}: |xy|=1 \}$: 
 $$
 x^{t+1}=
 \begin{cases}
 +, \text{ if } x^t=\circ \text{ and } \delta_x^t>0 \text{ and } \rho_x < \theta^t \text{ and } \gamma_x <c \text{ and } \beta_x <c\\
 \-,  \text{ if } x^t=+\\
 \#,  \text{ if } x^t=-\\
 \circ, \text{ otherwise }
 \end{cases}
 $$   
 where $\delta_x^t$ is a sum of excited neighbours, $\delta_x^t=\sum_{y \in u(x)}  \chi(y^t,+)$, where $\chi(y^t,+)=1$ if $y^t=+$, and 0 otherwise; $c=230$ and $\theta^t$ is updated by the rule
 $$
 \theta^{t+1}=
 \begin{cases}
 \theta^t - \delta \theta, \text{ if } S^t < \epsilon\\
  \theta^t + \delta \theta, \text{ if } S^t > \epsilon+d\\
  \theta^t,  \text{ otherwise }
 \end{cases}
 $$
 where $S^t = \sum_{y \in \mathbf{A}} \chi(y^t,+)$, $d$ is a range of excitation activity.   
 Value  $\epsilon$ represents a minimum level of activity or energy of imitated slime mould, and value $\epsilon+d$ is 
 a an upper limit of activity.   A resting cell excites if it has at last one excited neighbour ($\delta_x^t>0$) and red value of a pixel (of satellite image) corresponding to cell $x$ does not exceed $\theta^t$ and green and blue values do not exceed a constant $c$,  the constant $c$ is chosen 230, so excitation wave does not propagate outside the country's shape (Fig.~\ref{CA_US_POINTERS}ac). The automaton behaves like a classical excitable automaton with the only difference that a refractory state is not changed back to a resting state but to an absorbing precipitate state.
 
 Value of $\theta^t$ (the same for all cells) is updated at very step  $t$ of iteration in such a manner to keep excitation propagating yet minimise propagation over high elevation sites (high elevations sites have higher values of $\rho_x$). Namely, when number of excited sites  $S^t$ is less than a specified value $\epsilon$ the threshold $\theta^t$ is
 increased to let excitation propagate over higher elevations and do not extinct.  However, when there are too 
 many excited sites, $S^t > \epsilon+d$, the threshold $\theta^t$ is decreased and therefore excitation fronts 
 can enter less elevations. 
 
Excitation wave-fronts imitate active growing zones of plasmodium while pointers imitate protoplasmic pre-network. A thickest protoplasmic tube is selected by tracing a path along the pointers from destination site to start site. 
 
 Initially all cells are resting but a cell corresponding to start site (Flensburg in Germany and Newport in USA) is excited. Excitation wave-front propagate from the start site to destination site avoiding elevations specified by $\theta^t$ 
 (Fig.~\ref{CA_US_POINTERS}ac). Iterations are halted when either excitation reaches destination-site (F\"{u}ssen in Germany and Boston in USA) or becomes  extinct. 
 
 To trace a shortest path from a destination site to a source-site we must supply cells with unit vectors, or pointers to their neighbour, showing a local direction excitation came from. Each cell $x$ is assigned a pointer 
 $\mathbf{p}_x$ which has a zero length at the beginning of cellular automaton development. The pointer $p_x$ becomes non-zero vector when cell $x$ is excited:
 $$
 p_x^t=
 \begin{cases}
 \sum_{y \in u(x): y^{t-1}=-} \overline{xy}/\sum_{y \in u(x)} \chi(y^{t-1},-), \text{ if } x^t=+\\
 p_x^{t-1}, \text{ otherwise }
 \end{cases}
 $$
 Exemplar configurations of pointers are shown in Fig.~\ref{CA_US_POINTERS}bd.
 
 \begin{figure}[!tbp]
\centering
\subfigure[$\epsilon=30$]{\includegraphics[width=0.46\textwidth]{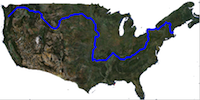}}
\subfigure[$\epsilon=50$]{\includegraphics[width=0.46\textwidth]{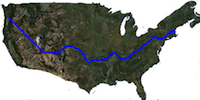}}
\subfigure[$\epsilon=60$]{\includegraphics[width=0.46\textwidth]{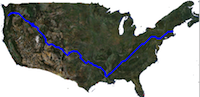}}
\subfigure[$\epsilon=70$]{\includegraphics[width=0.46\textwidth]{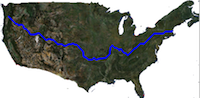}}
\subfigure[$\epsilon=80$]{\includegraphics[width=0.46\textwidth]{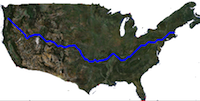}}
\subfigure[$\epsilon=90$]{\includegraphics[width=0.46\textwidth]{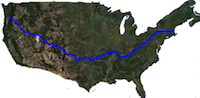}}
\subfigure[$\epsilon=100$]{\includegraphics[width=0.46\textwidth]{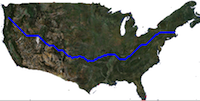}}
\subfigure[$\epsilon=110$]{\includegraphics[width=0.46\textwidth]{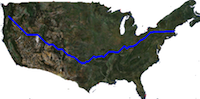}}
\subfigure[$\epsilon=140$]{\includegraphics[width=0.46\textwidth]{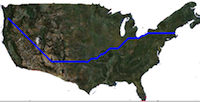}}
\subfigure[$\epsilon=150$]{\includegraphics[width=0.46\textwidth]{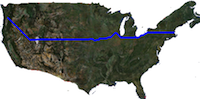}}
\caption{Examples of paths between Newport area and Boston area developed by cellular automaton on
a satellite image of USA. The following parameters are used: $\epsilon$ is indicated in subfigures,
$d=10$, $\theta^0=150$ and $\delta \theta=10$. Cellular automaton has 500$\times$253 cells.}
\label{CA_US_PATHS}
\end{figure}

\begin{figure}[!tbp]
\centering
\subfigure[$\theta^0=150$, $\epsilon=20$, $d=90$]{\includegraphics[width=0.3\textwidth]{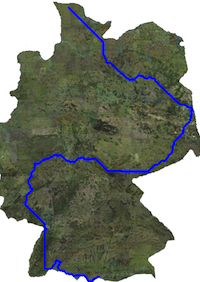}}
\subfigure[$\theta^0=150$, $\epsilon=65$, $d=90$]{\includegraphics[width=0.3\textwidth]{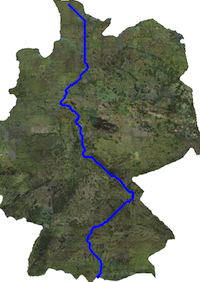}}
\subfigure[$\theta^0=150$, $\epsilon=110$, $d=90$]{\includegraphics[width=0.3\textwidth]{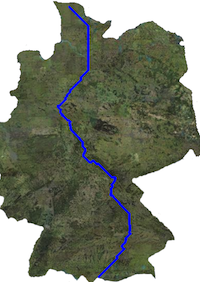}}
\subfigure[$\theta^0=200$, $\epsilon=100$, $d=100$]{\includegraphics[width=0.3\textwidth]{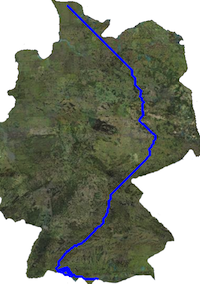}}
\subfigure[$\theta^0=200$, $\epsilon=130$, $d=100$]{\includegraphics[width=0.3\textwidth]{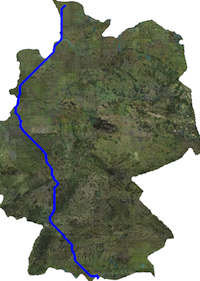}}
\subfigure[$\theta^0=200$, $\epsilon=160$, $d=100$]{\includegraphics[width=0.3\textwidth]{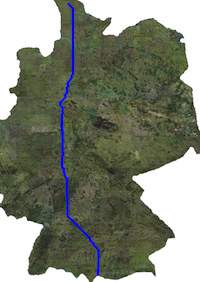}}
\caption{Examples of paths between Flensburg area and F\"{u}ssen area developed by cellular automaton on
a satellite image of Germany. The following parameters are used: $\theta^0$, $\epsilon$ and $d$ are indicated in
the subfigures $\delta \theta=15$. Cellular automaton has 505$\times$707 cells.}
\label{CA_GERMANY_PATHS}
\end{figure}

When we want to trace a shortest path --- as presented by configuration of pointers developed in an an excitable medium --- we start with destination site 
and then follow pointers to extract sites of the path one by one. In rare situations when path becomes loop locked we 
we choose next neighbour to move to in random and continue following pointers (Figs.~\ref{CA_US_PATHS} and \ref{CA_GERMANY_PATHS}).

\begin{figure}
\centering
\subfigure[Route length]{\includegraphics[width=0.8\textwidth]{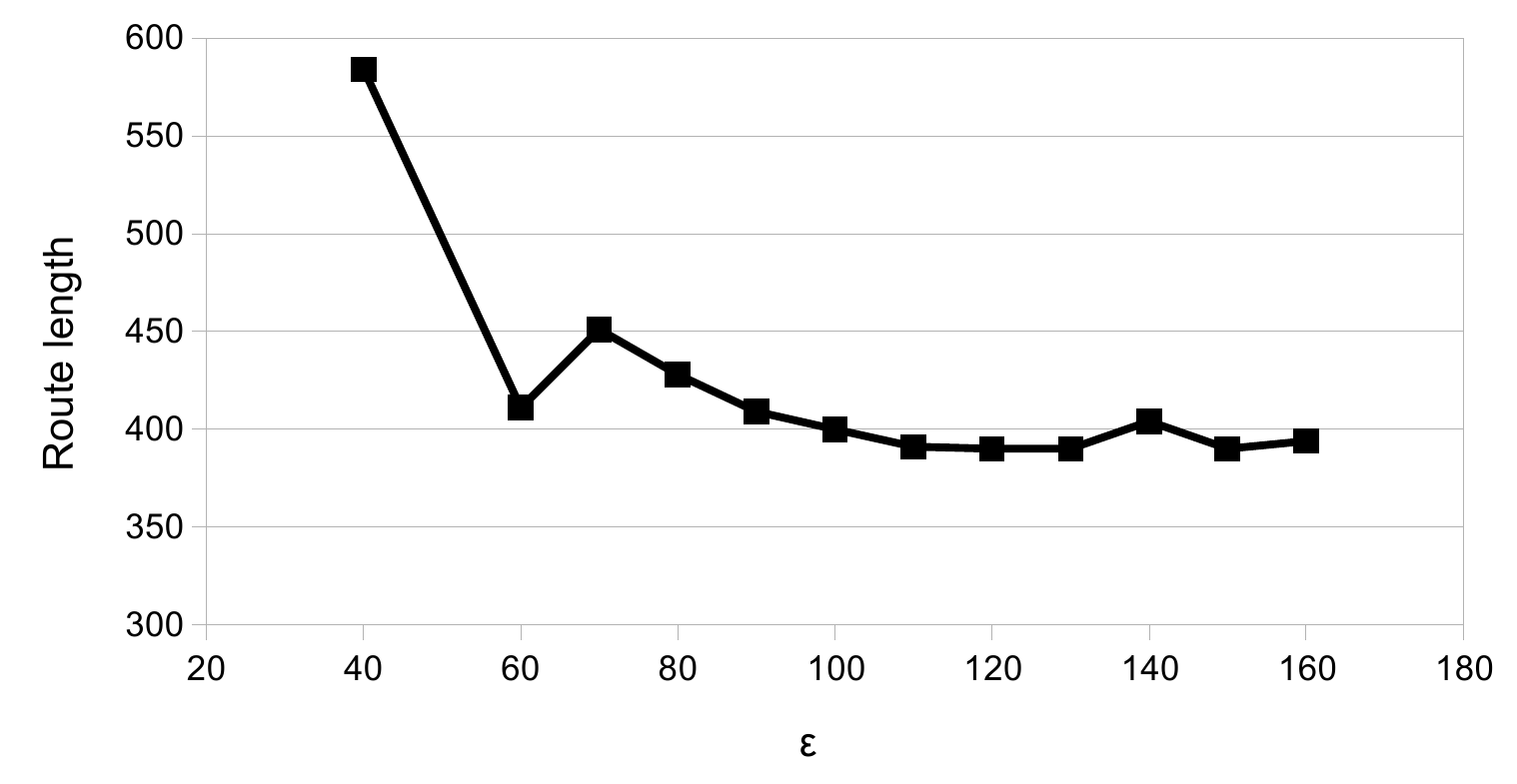}}
\subfigure[Number of excited sites]{\includegraphics[width=0.8\textwidth]{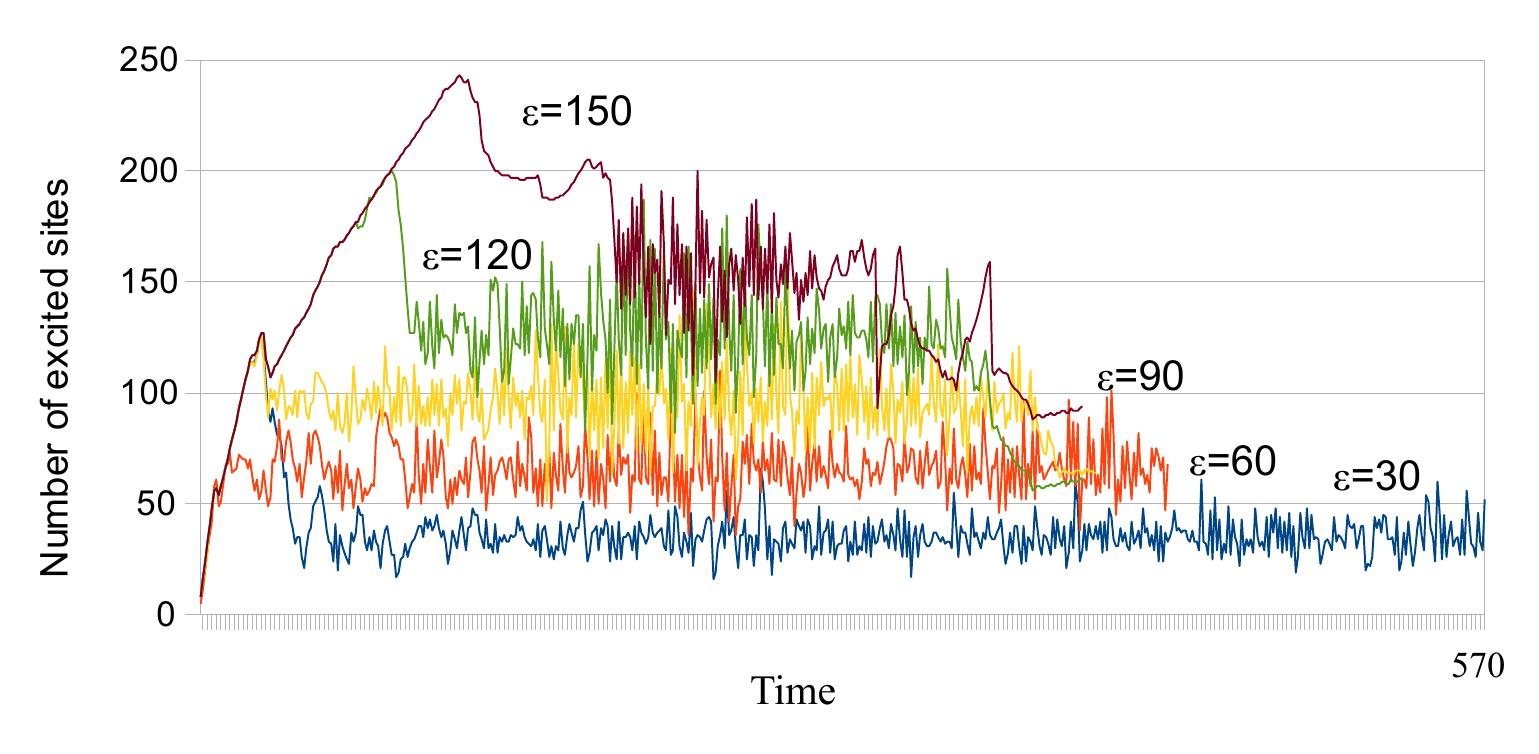}}
\caption{Some integral characteristics of the model, illustrated in Fig.~\ref{CA_US_PATHS}. (a)~Lengths of routes approximated. (b)~Dynamics of excitation level. The following parameters are used:  $d=10$, 
$\theta^0=150$,  and  $\delta \theta=10$.}
\label{CA_USA_STATS}
\end{figure}

\begin{finding}
A length of a path developed by the excitable cellular automaton decreases with increase of $\epsilon$.
\end{finding}

Value $\theta^t$ is increased when a level of excitation (number of excited cells) exceeds $\epsilon + d$. Thus the larger is $\epsilon$ the less chances for $\theta^t$ to increase. Therefore, for large values of $\epsilon$ 
more sites can be excited, and thus excitation wave fronts can climb on higher elevations. The dependency is illustrated in Fig.~\ref{CA_USA_STATS}a. Lengths of paths calculated by the automaton drops substantially when $\epsilon$ increases from  40 to  60, and then stabilises when $\epsilon$ exceeds 100 (Fig.~\ref{CA_USA_STATS}a). A path becomes shorter but at the costs of more energy spent on ascending elevations. 

Excitation propagates as a single wave-front only at the beginning of development, see high rise in a number of excited sites in Fig.~\ref{CA_USA_STATS}b. As soon as excitation level exceeds $\epsilon+d$ value of $\theta^t$ decreases and thus less elevations can be climbed on. This leads to splitting of a single excitation wave-front to several fragments of wave-fronts whose size and configuration are controlled by a configuration of elevations in the fronts' vicinity. Thus for a low values of $\epsilon$ we have many small wave-fronts scouting the satellite image  (Fig.~\ref{CA_USA_STATS}a). For high values of $\epsilon$  we observe a rather classical wave, which sometimes gets split by elevations (Fig.~\ref{CA_USA_STATS}c).

Figure~\ref{CA_US_PATHS} shows how topology of a path between Newport and Boston changes with increase of $\epsilon$ from 30 to 150. For almost any path in (Fig.~\ref{CA_US_PATHS}) we can find a similar protoplasmic route obtained in experiments with slime mould (Fig.~\ref{USAGermanyResults}c), especially for routes lying along or south of route 20. Paths developed by propagating wave-fronts for high values of $\epsilon$ reasonably close match route 20, 
see e.g. Fig.~\ref{CA_US_PATHS}j. 

Dependence of a path length on activity level is demonstrated with cellular automaton developing path between 
Flensburg area and F\"{u}ssen area in Germany (Fig.~\ref{CA_GERMANY_PATHS}).  For low levels of activity, e.g. 
$\epsilon=20$ (Fig.~\ref{CA_GERMANY_PATHS}a) path propagates  from Flensburg east to Brandenburg, enters Sachsen, propagates west-south-west along Erzgebirge, reflected slightly north by Th\"{u}ringen Wald. The path navigate between Kellerwald, Vogelsberg, Taunus and Spessart, propagates south along Shwarzwald, and only on reaching south boundary of Germany turns towards its destination site. The path matches several plasmodial routes 
obtained in laboratory experiments (Fig.~\ref{USAGermanyResults}d). 

When we increase $\epsilon$ to 65 a path developed by excitable cellular automaton propagates from 
Schleswig-Holstein through Hannover to Bremen and Niedersachsen and navigates around Harz. The path 
turns east to Th\"{u}ringen, gets reflected by north part of B\"{o}hmer-Wald to south-west and then propagates to its destination-site  (Fig.~\ref{CA_GERMANY_PATHS}b). For high values of $\epsilon$ path develops almost as a straight line   (Fig.~\ref{CA_GERMANY_PATHS}i).

In summary, plasmodium of \emph{P. polycephalum} does not compute a shortest path \emph{per se}, but the path optimal for the amount of resources involved. 

\section{Scoping experiments with 3D terrains of United Kingdom and Russia}
\label{UKRussia}

\begin{figure}[!tbp]
\centering
\subfigure[]{\includegraphics[width=0.35\textwidth]{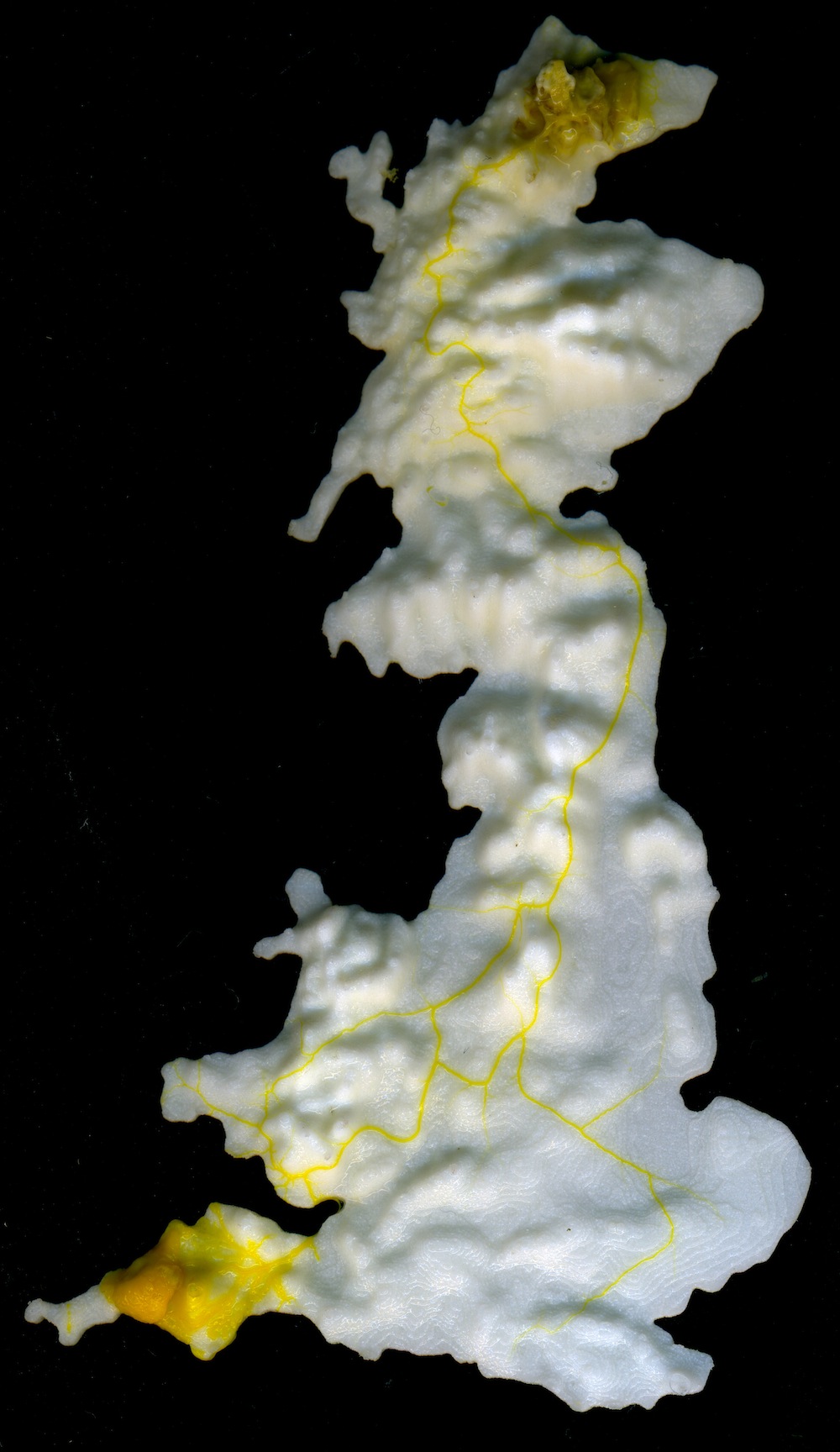}}
\subfigure[]{\includegraphics[width=0.5\textwidth]{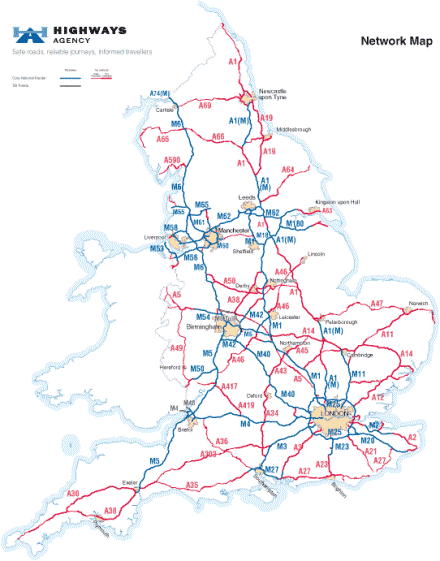}}
\subfigure[]{\includegraphics[width=0.35\textwidth]{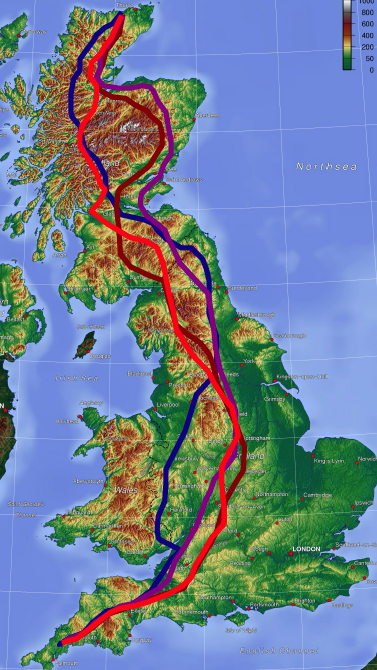}}
\caption{Slime mould on 3D terrain of United Kingdom. (a)~Scanned image of one experiment.
(b)~A scheme of major roads networks in UK~(b)~\cite{mapUK}}
\label{UK}
\end{figure}

Example of one of the scoping experiments with 3D terrain of the United Kingdom is illustrated in Fig.~\ref{UK}. We used
3D terrain  6.2 cm wide, 11.0 cm long with maximum elevation 1.0 cm. 
We inoculated plasmodium in the northern part of Scotland, in the area between Tongue and Thurso, and placed an 
oat flake (which acts as a chemoattractant) to western part of England, in the area between Truro and Penzance (due to size of an average oat flake, c. 5 mm, it was impossible to place the flake in the location of the Land's End). A network
of protoplasmic tubes developed does not match any major roads in the Highlands. Some matches between the roads and the tubes are observed when the slime mould approach Edinburgh in proximity of A84 and M9 (compare
Fig.~\ref{UK}a and b). South of Edinburgh the slime mould propagates along A1 and A697, passing close to Eyemouth. 

While propagating further south the plasmodium passes Durham, near Stockton-on-Tees, Leeds and Bradford area, Bolton and Crewe. On entering Wales the slime mould branches: one branch goes through Brecon to Carmarthen, another to the area of Cardiff and Newport (Fig.~\ref{UK}). The plasmodium crosses from Newport to Bristol, passes Bridgewater and then follows along M5. After passing Exeter the slime mould spreads all over Cornwall.   Also, in the area between Coventry and Northampton a redundant branch appears. It follows towards location of London along M1 and partly A38 and M42, and then follows south between A3 and M3 (Fig.~\ref{UK}).

Experimental image Fig.~\ref{UK}a shows well that while propagating from Scottish Highlands to Cornwall slime mould
avoids entering or climbing on top of major elevations. Namely, the plasmodium navigates around Ben Nevis and Grampian Mountains, avoids Southern Uplands by passing close to Firth of Forth. It passes east of Cheviots Hills and crosses Pennines in the least elevated place, as if along a line connecting York to Manchester. The slime mould avoids Cotswold Hills by going from Newport area to Bristol. Its redundant branch passes west of Chiltern Hills while propagating to London.

The experiment described above and illustrated in  (Fig.~\ref{UK}a) is quite an illustrative one. As you can see in the 
scheme  (Fig.~\ref{UK}c) of major protoplasmic tubes, extracted from images obtained in four experiments, the slime mould, in general, follows the same route as described above with minor variations in navigating around mountains.

\begin{figure}[!tbp]
\centering
\subfigure[]{\includegraphics[width=0.49\textwidth]{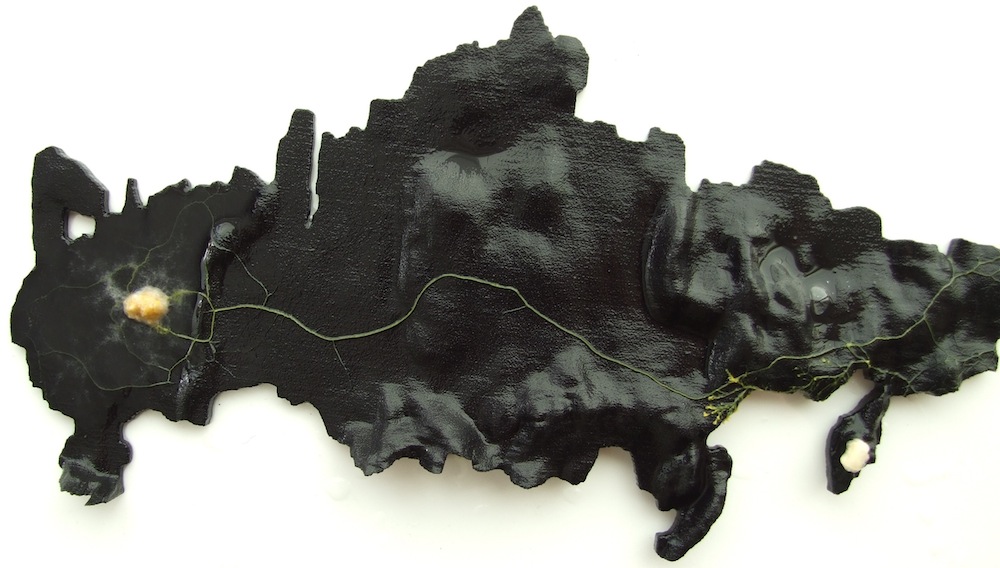}}
\subfigure[]{\includegraphics[width=0.49\textwidth]{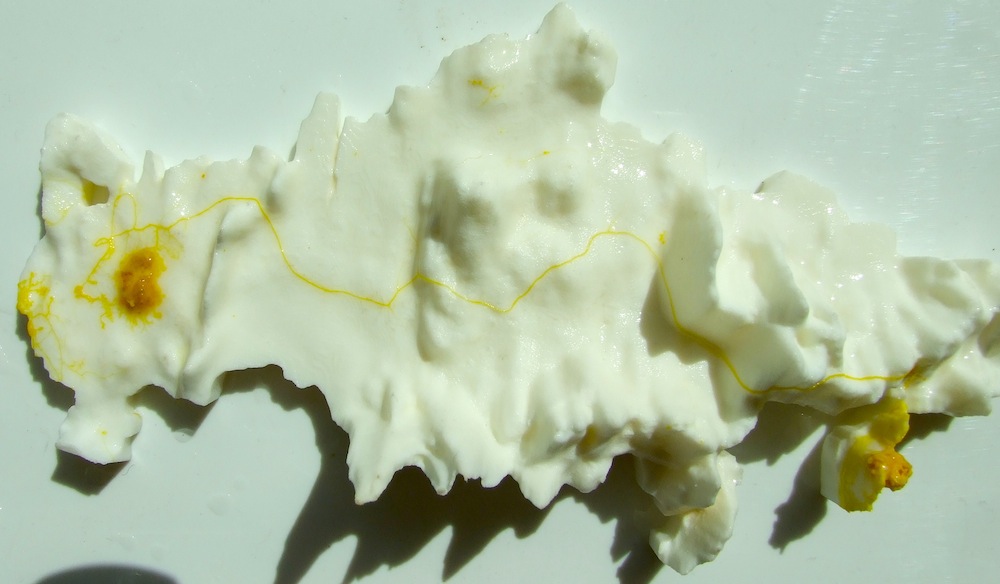}}
\subfigure[]{\includegraphics[width=0.49\textwidth]{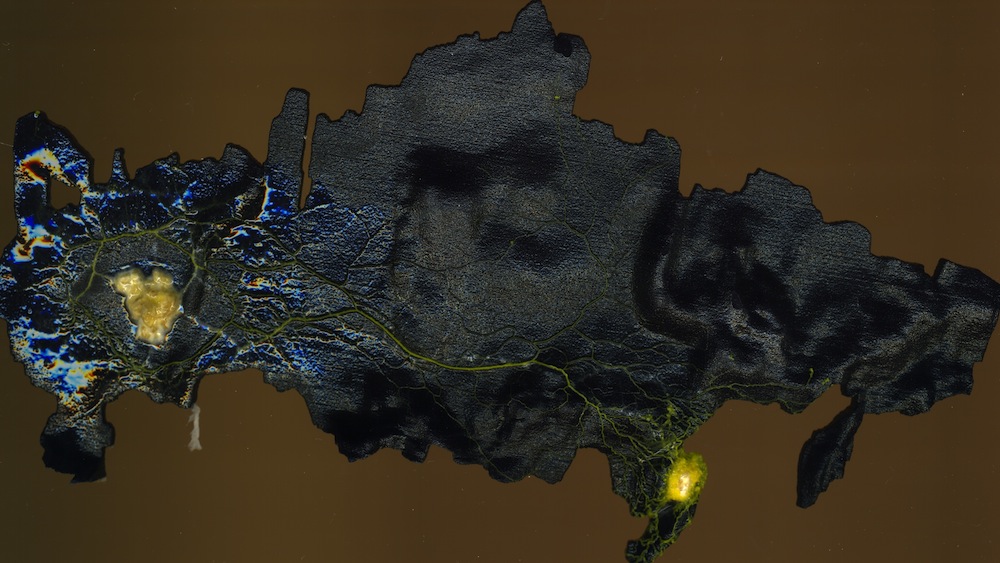}}
\subfigure[]{\includegraphics[width=0.49\textwidth]{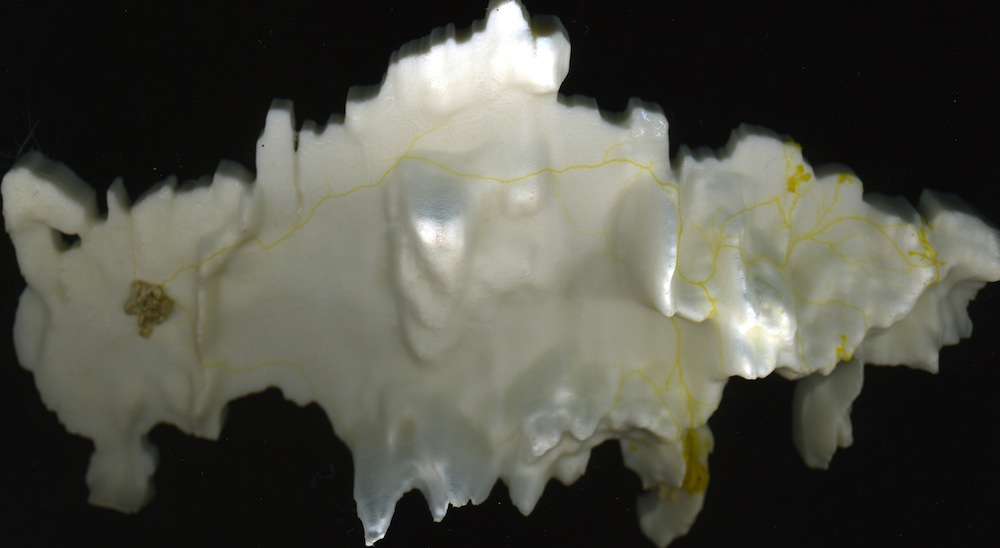}}
\subfigure[]{\includegraphics[width=0.9\textwidth]{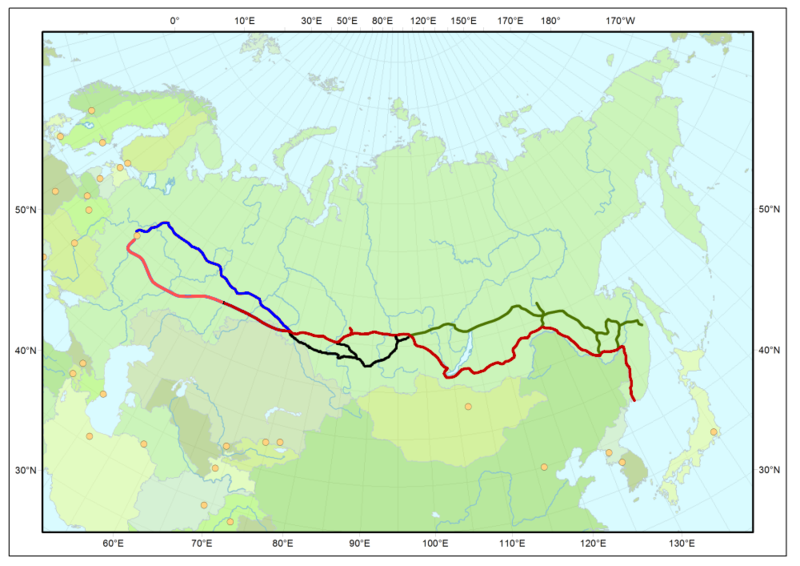}}
\caption{Images of four experimental setups, where slime mould develops a transport network 
between (ab)~Moscow and Petropavlovsk-Kamchatsky and (cd)~Moscow and Vladivostok.
(e)~Map of Trans-Siberian railway (red),  Baikal-Amur mainline (green)~\cite{kuhn}.}
\label{RussiaExperiments}
\end{figure}

\begin{figure}[!tbp]
\centering
\subfigure[]{\includegraphics[width=0.47\textwidth]{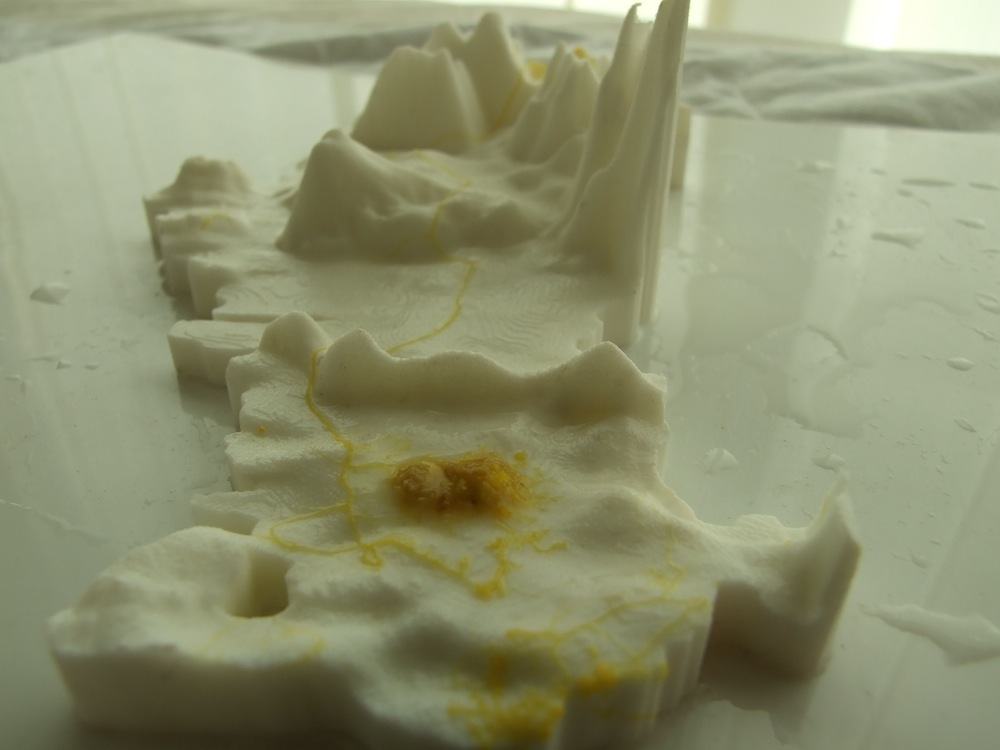}}
\subfigure[]{\includegraphics[width=0.47\textwidth]{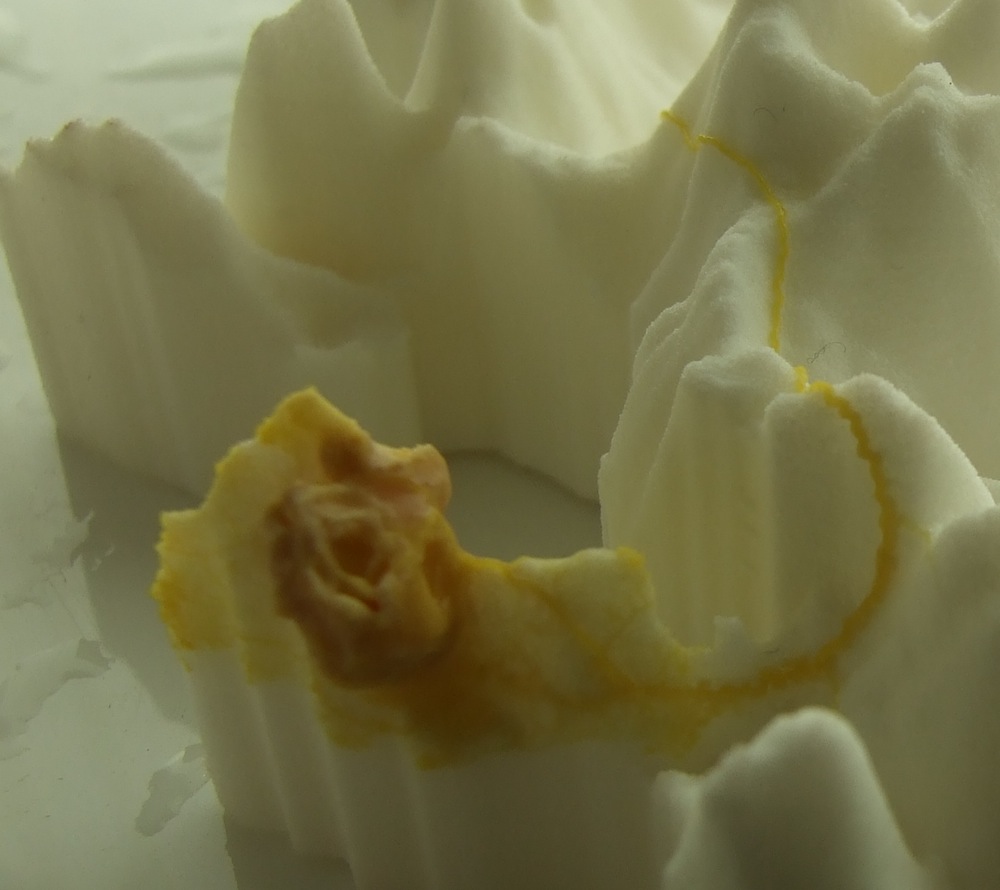}}
\subfigure[]{\includegraphics[width=0.75\textwidth]{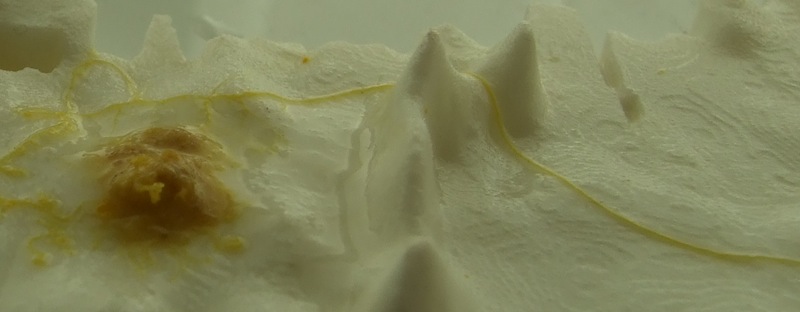}}
\subfigure[]{\includegraphics[width=0.75\textwidth]{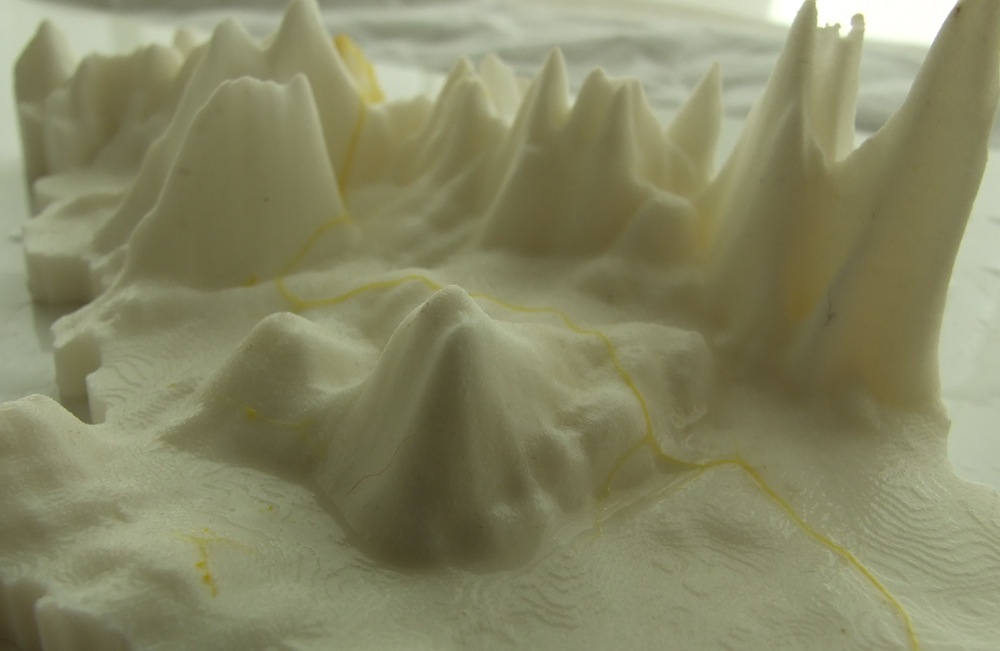}}
\caption{Example of slime mould navigating mountains.
 (a)~West to east view of slime mould crossing Ural Mountains along a virtual line connecting Vorkuta to 
 Labytnangi.
 (b)~Slime mould propagating along slops of Kamchatka mountains and into Kamchatka peninsula. 
 (c)~South to north view of slime mould crossing Ural Mountains along a virtual line connecting Vorkuta to 
 Labytnangi.
 (d)~Slime mould passes through Central Siberian Plateau north of Enashimsky mountain, in the region of 
 Tura city.}
\label{passingmoutainsRussia}
\end{figure}

In experiments with Russia two terrains were used. One terrain, white, is 20.0 cm wide,  10.6 long and 5.0 cm high (at highest elevation, peaks of Elbrus were cut of), another,  black in colour,  has the same length and width yet, highest elevation is 1.5 cm.  Plasmodium was inoculated in a domain between Moscow and Kazan and oat flakes placed in 
locations of Petropavlovsk-Kamchatsky or Vladivostok (Fig.~\ref{RussiaExperiments}). 

Slime mould navigated towards sources of chemo-attractants often passing mountains in the places of lowest 
elevations (Fig.~\ref{passingmoutainsRussia}) and demonstrated a high-degree of variability in developed protoplasmic transport links. For example, plasmodium growing on a higher-elevated model (white terrain) tended to cross Ural Mountains in their northern parts  (Fig.~\ref{RussiaExperiments}bd) while the plasmodium growing on lower-elevated model (black terrain) avoided Ural Mountains in their southern parts  (Fig.~\ref{RussiaExperiments}ac). 
 
 Only in one of the experiments illustrated the slime mould avoided  Central Siberian Plateau by passing around its northern terminus (Fig.~\ref{RussiaExperiments}d) . In three other experiments the plasmodium developed a protoplasmic link which either avoids the elevation by passing between Central Siberian Plateau and Sayan 
 Mountains (Fig.~\ref{RussiaExperiments}c) or crossing the Plateau along Podkamennaya Tunguska river (Fig.~\ref{RussiaExperiments}ab).

Few experiments undertaken with 3D terrain of Russia  did not provide a substantial evidence that plasmodium's transport networks connecting Moscow region and Vladivostok closely match Trans-Siberian railway or Baikal-Amur mainline (Fig.~\ref{RussiaExperiments}e). One experiment, shown in Fig.~\ref{RussiaExperiments}c, 
was a pleasant surprise: major protoplasmic tube grown along Trans-Siberian railway from Moscow-Kazan region to Omsk
and then to Krasnoyarsk. The protoplasmic tube then propagates along Baikal-Amur mainline till turning south towards Vladivostok.

\section{Discussion}
\label{discussion}

Acellular slime mould \emph{Physarum polycephalum} approximates shortest path in a labyrinth when pieces 
of plasmodium are distributed in all channels of the labyrinth~\cite{nakagaki_2001}. The slime mould also finds 
a shortest path in a labyrinth from a single start site while being assisted with diffusion gradient of 
chemo-attractants~\cite{adamatzky_maze}. The slime mould satisfactory approximates man-made transport networks between major urban areas when these areas are represented by sources of nutrients~\cite{adamatzky_bioevaluation}.
All the above experiments  were conducted on a flat agar gel. A few questions remained. How well the slime mould approximates a path on a substrate with elevations? Do slime mould  paths grown on flat substrate differ from the paths developed on 3D models of countries? 

To find an answer we conducted a series of experimental laboratory studies on approximation of route 20, the longest road in USA, and autobahn 7, the longest national motorway in Europe on 3D terrain plastic models of USA and Germany.

We inoculated plasmodium of  \emph{P. polycephalum} at one end-point of a road approximated and placed sources of nutrients at another end-point. The slime mould grown from its inoculation site, explored 3D terrains, reached sources of nutrients and represented a path between start site and destination site with its sickest protoplasmic tube.  We have also done control experiments on a flat agar gel, and on substrates without sources of nutrients.  

In laboratory experiments we demonstrated that route 20 is approximated by live slime mould on 3D terrain closer than autobahn 7. Autobahn 7 is better approximated on a flat agar gel.  On flat surface average protoplasmic route in Germany is just 1.005 times longer than autobahn 7 while an average protoplasmic tube in USA is just 1.046 times longer than route 20. On 3D terrains of USA average slime route is 1.095 times longer than route 20, while on terrain of Germany average slime route is 1.158 times longer. It is not yet clear why plasmodium of \emph{P. polycephalum} performs better on USA terrain. We can speculate this is due to location of transport routes approximates relative to national boundaries of the countries.   Route 20 in USA leans towards north boundary of the country while autobahn 7 lies more or less equidistantly from east and west boundaries of Germany. Therefore in USA slime mould can substantially deviate from route 20 by propagating only south, in Germany growing slime mould can deviate west and east. 

In computer simulations we imitated active growing zones of plasmodium with excitation wave-fronts and structure of protoplasmic tubes with pointers in excitable cellular automata. We shown that configuration of a path developed is determined by a level of activity of an imitated plasmodium, expressed via relation between minimum and maximum allowed levels of excitation  and maximum heigh of elevations allowed to climb on. Higher levels of activity lead to shorter paths developed.

\end{document}